\documentclass[a4paper,11pt,UKenglish]{article}

\usepackage{amsmath,amssymb, xcolor, amsthm}
\usepackage{mathtools}
\usepackage{url}
\usepackage{hyperref}
\usepackage{fullpage}
\usepackage{shuffle}
\usepackage{wasysym}
\usepackage{ulem}
\usepackage{pstricks}
\usepackage{pst-plot, pst-node, pst-text, pst-tree}
\usepackage{rotating} 

\usepackage{tikz}

\usepackage[linesnumbered,boxed,vlined]{algorithm2e}	
\SetAlCapSkip{0.5em}

\usepackage{graphicx}
\usepackage{indentfirst}
\usepackage[capitalize]{cleveref}

\usepackage{todonotes}

\hypersetup{colorlinks=true, linkcolor=black, citecolor=blue, urlcolor=blue}
\theoremstyle{definition}

\theoremstyle{plain}
\newtheorem{theorem}{Theorem}[section]
\newtheorem{prop}[theorem]{Proposition}
\newtheorem{lemma}[theorem]{Lemma}

\newtheorem*{conj}{Conjecture}

\DeclareMathOperator{\bub}{\mathbf{B}}
\DeclareMathOperator{\que}{\mathbf{Q}}

\DeclareMathOperator{\psb}{\mathbf{PSB}}

\DeclareMathOperator{\stack}{\mathbf{S}}

\newcommand{\Av}{\mathrm{Av}}

\date{}

\author{Lapo Cioni\thanks{Dipartimento di Statistica, Informatica e Applicazioni, Universit\'a degli Studi di Firenze, Firenze, Italy. {\tt lapo.cioni@unifi.it}} \and Luca Ferrari\thanks{Dipartimento di Matematica e Informatica ``U. Dini'',
Universit\`a degli Studi di Firenze, Firenze, Italy. {
\tt\ luca.ferrari@unifi.it} Member of the INdAM research group GNCS; partially supported by the 2024 INdAM-GNCS project CUP\_E53C23001670001} \and Rebecca Smith\thanks{Department of Mathematics, SUNY Brockport, Brockport, New York, USA. {\tt rnsmith@brockport.edu}}}

\title{Sorting permutations using a pop stack with a bypass}

\begin{document}

\maketitle

\def\sdwys #1{\xHyphenate#1$\wholeString}
\def\xHyphenate#1#2\wholeString {\if#1$%
\else\say{\ensuremath{#1}}\hspace{2pt}%
\takeTheRest#2\ofTheString
\fi}
\def\takeTheRest#1\ofTheString\fi
{\fi \xHyphenate#1\wholeString}
\def\say#1{\begin{turn}{-90}\ensuremath{#1}\end{turn}}

\newenvironment{onestack}
{
	\begin{scriptsize}
	\psset{xunit=0.0355in, yunit=0.0355in, linewidth=0.02in}
	\begin{pspicture}(0,-2)(32,20)
	\psline{c-c}(0,15)(10,15)
	\psline{c-c}(13,15)(13,2)(19,2)(19,15)
	\psline{c-c}(22,15)(32,15)
	\rput[l](-0.5,12.5){\mbox{output}}
	\rput[r](32,12.5){\mbox{input}}
}
{
	\end{pspicture}
	\end{scriptsize}
}

\newcommand{\fillstack}[3]{%
	\rput[l](-0.5,17.5){\ensuremath{#1}}
	\rput[c](16.1, 7.5){\begin{sideways}{\sdwys{#2}}\end{sideways}}
	\rput[r](32,17.5){\ensuremath{#3}}
}

\newcommand{\op}[1]{%
    \rput[c](16,-0.5){#1}
}

\newenvironment{twostack}
{
	\begin{scriptsize}
	\psset{xunit=0.0355in, yunit=0.0355in, linewidth=0.02in}
	\begin{pspicture}(0,-2)(50,40)
	\psline{c-c}(0,17.5)(10,17.5)
	\psline{c-c}(13,33)(13,20)(19,20)(19,33)
    \psline{c-c}(13,15)(13,2)(19,2)(19,15)
	\psline{c-c}(22,17.5)(49,17.5)
	\rput[l](-0.5,15.5){\mbox{output}}
	\rput[r](40,15.5){\mbox{input}}
}
{
	\end{pspicture}
	\end{scriptsize}
}

\newcommand{\fillstacktwo}[4]{%
	\rput[l](-0.5,20.5){\ensuremath{#1}}
	\rput[c](16.1, 25.5){\begin{sideways}{\sdwys{#2}}\end{sideways}}
	\rput[c](16.1, 7.5){\begin{sideways}{\sdwys{#3}}\end{sideways}}
	\rput[r](49,20.5){\ensuremath{#4}}
}

\newcommand{\optwo}[1]{%
    \rput[c](16,-0.5){#1}
}

\begin{abstract}
We introduce a new sorting device for permutations which makes use of a pop stack augmented with a bypass operation. This results in a sorting machine, which is more powerful than the usual Popstacksort algorithm and seems to have never been investigated previously.

In the present paper, we give a characterization of sortable permutations in terms of forbidden patterns and reinterpret the resulting enumerating sequence using a class of restricted Motzkin paths. Moreover, we describe an algorithm to compute the set of all preimages of a given permutation, thanks to which we characterize permutations having a small number of preimages. Finally, we provide a full description of the preimages of principal classes of permutations, and we discuss the device consisting of two pop stacks in parallel, again with a bypass operation.   
\end{abstract}

\section{Introduction}

Sorting disciplines for permutations constitute a flourishing research area in contemporary combinatorics. Starting from the stacksorting procedure introduced by Knuth \cite{K}, dozens of articles (many of which are very recent) have explored the subject. For example, machines making use of various types of containers have been studied, as well as networks of the corresponding devices. Typical containers that are considered in this context include stacks, queues, and deques, as well as their ``pop" versions. In particular, a \emph{pop stack} is a stack whose push and pop operations are similar to the usual ones for stacks. The distinction is that a pop operation extracts \emph{all} the elements from the pop stack, rather than just the element on the top. The sorting power of a pop stack was originally studied by Avis and Newborn \cite{AN}, who also considered pop stacks in parallel. Concerning pop stacks in series, we mention that the right-greedy version of pop stacks in series was introduced by Pudwell and Smith \cite{PS} and was successively studied also by Claesson and Gu\dh{}mundsson \cite{CG}.

\bigskip

When dealing with a classical stack, the algorithm \texttt{Stacksort} introduced by Knuth uses only two operations, namely the push operation (which pushes the current element of the input into the stack) and the pop operation (which pops the stack by moving the top of the stack into the output). Recall that permutations sortable using \texttt{Stacksort} are characterized as those avoiding the pattern 231. 
In this case, a bypass operation, which sends the current element of the input directly into the output, is clearly redundant, since it can be simulated by means of a push followed by a pop. This is clearly false if we replace the stack with a queue: without a bypass operation, the resulting device would be able to sort only the identity permutations, whereas adding a bypass gives rise to a much more interesting device. In the latter case, an optimal sorting algorithm, called \texttt{Queuesort}, is described in \cite{M} and further studied in \cite{CF1}, and sortable permutations can be characterized as those avoiding the pattern 321.

When we move to the ``pop" versions, again in the case of a pop queue a bypass operation is necessary in order to have nontrivial algorithms \cite{CF2}. On the other hand, in the case of a pop stack, having or not having a bypass operation gives rise to two different (and nontrivial) sorting devices (unlike what happens for a classical stack). Curiously, there is no evidence in the literature of papers investigating the sorting properties of a pop stack with a bypass. This paper aims to fill this gap. Clearly, adding a bypass to a pop stack slightly increases the sorting power of the device and poses interesting combinatorial questions on its properties, which we analyze in some detail. More specifically, the content of our paper is the following.

In Section \ref{sortable} we characterize the set of sortable permutations in terms of two forbidden patterns; the enumeration of the resulting class of pattern avoiding permutations was already known (odd-indexed Fibonacci numbers, sequence A001519 in \cite{Sl}); however, in Section \ref{motzkin_link} we give an independent proof of this enumerative result by describing a bijective link with a restricted class of Motzkin paths. In Section \ref{preimages_perms} we describe an algorithm to compute the preimage of a given permutation and use it to characterize and enumerate permutations having 0,1, and 2 preimages. Section \ref{classes} contains a complete description of the preimages of principal classes of permutations, by determining in which cases we obtain classes (and in such cases by determining the basis of the resulting classes). Section \ref{composition} is devoted to the characterization and (partly) enumeration of
the sets of sortable permutations for the compositions of our sorting algorithm with other classical sorting algorithms (the characterization being given in terms of forbidden patterns). Finally, in Section \ref{parallel} we consider the device consisting of two pop stacks in parallel with a bypass, and we determine the basis of the associated set of sortable permutations, and in Section \ref{final} we give some hints for further work.

\section{Preliminaries on permutations and sorting algorithms}

Let $[1,n]=\{1,\dots,n\}$. A permutation $\pi$ of size $n$ is a bijection from $[1,n]$ to $[1,n]$. It can be represented in linear form as $\pi=\pi_1\cdots\pi_n$, where $\pi_i=\pi(i)$, for all $1\le i\le n$. The \emph{identity permutation} of size $n$ is $12\cdots n$, and is denoted $id_n$.
We denote by $S_n$ the set of all permutations of size $n$ and by $S=\bigcup_{n\in \mathbb{N}}S_n$ the set of all permutations.

For a given permutation $\pi$, we say that two elements are \emph{adjacent} when their positions are consecutive integers, whereas we say that they are \emph{consecutive} when their values are consecutive integers. For instance, if $\pi =3645712$, the elements 1 and 7 are adjacent, the elements 5 and 6 are consecutive, and the elements 4 and 5 are both adjacent and consecutive. We warn the reader that we will strictly adhere to this terminology regarding the terms ``consecutive" and ``adjacent" throughout the paper.

Given $\pi =\pi_1 \cdots \pi_n \in S_n$, an element $\pi_i$ is called a \emph{left-to-right maximum} when it is larger than all the elements to its left (that is, $\pi_i =\max \{ \pi_j \, |\, j\leq i\}$). We denote by $LTR(\pi)$ the set of left-to-right maxima of $\pi$. Therefore, the set of left-to-right maxima of the above permutation is $LTR(3645712)=\{ 3,6,7\}$. Moreover, the \emph{reverse} of $\pi$ is the permutation $\pi^{r}=\pi_n \cdots \pi_1$.

Given $\pi \in S_n$ and $\tau =\tau_1 \cdots \tau_m \in S_m$, the \emph{direct sum} of $\pi$ and $\tau$ is the permutation $\pi \oplus \tau$ obtained by concatenating $\pi$ with the sequence $(\tau_1 +n)\cdots (\tau_m +n)$ (obtained from $\tau$ by adding $n$ to each of its elements). For instance, the direct sum of the permutations $3142$ and $42315$ is $314286759$.

We note that any sequence of $n$ distinct integers can be regarded as a permutation by appropriately rescaling its elements to $[1,n]$. Indeed, two sequences are said to be \emph{order-isomorphic} if their elements are in the same relative order. This notion is useful in the definition of a pattern that we give next.

Let $\pi$, $\rho$ be two permutations of size $n$ and $k$, respectively. Then $\rho$ is a \emph{pattern} of $\pi$ (or $\pi$ contains the pattern $\rho$) when there exists a subsequence $\pi_{i_1}\cdots\pi_{i_m}$ of $\pi$, with $i_1\le \dots\le i_m$, which is order-isomorphic to $\rho$. Such a subsequence is also called an \emph{occurrence} of the pattern $\rho$.
Conversely, if $\pi$ does not contain $\rho$, then we say that $\pi$ \emph{avoids} $\rho$. The set of all permutations avoiding a pattern $\rho$ is denoted $\Av (\rho )$, and similarly, for $T\subseteq S$, we write $\Av (T)$ for the set of permutations avoiding \emph{each} pattern of $T$. We also use the notations $\Av_n (\rho )$ and $\Av_n (T)$ to denote the sets of permutations of size $n$ avoiding $\rho$ and $T$, respectively. As an example, the permutation $35142$ contains the patterns $213$ and $321$ (among others), but avoids the pattern $123$. In particular, an occurrence of the pattern 21 is called an \emph{inversion}.

The notion of pattern defines a partial order $\leq$ on $S$, by declaring $\rho\le\pi$ whenever $\rho$ is a pattern of $\pi$.
The resulting poset is called the \emph{permutation pattern poset}, and its down-sets\footnote{A \emph{down-set} of a poset $P$ is a subset of $P$ which is closed downwards; dually, an \emph{up-set} is a subset of $P$ which is closed upward.} are usually called \emph{classes}. It is not difficult to see that any nonempty class of the permutation pattern poset can be defined in terms of the avoidance of a set of patterns. More formally, given nonempty $\mathcal{D}\subseteq S$, $\mathcal{D}$ is a class if and only if there exists an antichain $T\subseteq S$ such that $\mathcal{D}=\Av (T)$. In this case, we say that $T$ is the \emph{basis} of the class $\mathcal{D}$.

There are also several variations of the above notion of pattern. One that is relevant for our paper (specifically in Proposition \ref{stackPopstack}) is that of a barred pattern. A \emph{barred pattern} is a permutation in which some of the entries are barred. For a permutation $\pi$ to avoid the barred pattern $\rho$ means that every set of entries of $\pi$ which form a copy of the nonbarred entries of $\rho$ can be extended to form a copy of all entries of $\rho$. For example, the permutation $462351$ avoids the barred pattern $3\bar{5}241$, as it has two occurrences of 3241 (namely $4251$ and $4351$), which can both be extended to an occurrence of $35241$.

To conclude this section on preliminaries, we briefly recall a few sorting algorithms for permutations that will be used in our paper. We will not give formal definitions of the algorithms, instead settling for an informal description of each of them.

The classical \texttt{Stacksort} algorithm was first introduced by Knuth \cite{K}, and tries to sort a permutation by making use of a stack as follows: the entries of the input permutation are processed from left to right and, when the current entry is smaller than the top of the stack (or the stack is empty), then such entry is pushed into the stack; otherwise the stack is popped, and the top of the stack is moved in the next available position of the output (i.e. to the immediate right of the elements already in the output). When all entries for the input permutations have been processed, the content of the stack is poured into the output. As is well known, the set of sortable permutations for \texttt{Stacksort} is the set of 231-avoiding permutations, which are counted by Catalan numbers. A similar sorting algorithm is \texttt{Queuesort} \cite{M}, where the stack is replaced with a queue and a bypass operation is also allowed. In this case, the algorithm proceeds by inserting the current entry into the queue if it is larger than the element at the back of the queue; otherwise, the current entry is compared with the element in front of the queue, and the smaller one is output. The set of sortable permutations using \texttt{Queuesort} is known to be the set of 321-avoiding permutations, which is again counted by Catalan numbers. Finally, we mention the very classical sorting algorithm \texttt{Bubblesort}, whose description can be found in most standard algorithm textbooks. In fact, our version of \texttt{Bubblesort} consists of a single pass of the algorithm reported in textbooks. In terms of \texttt{Stacksort} and \texttt{Queuesort}, \texttt{Bubblesort} can be interpreted as a special case of any of the two in which the container has depth 1. Permutations sortable through \texttt{Bubblesort} are precisely those that avoid the two patterns 231 and 321, and are counted by the sequence of the powers of two \cite{AABCD}.

\section{An optimal algorithm to sort permutations using a pop stack with a bypass}\label{sortable}

A \emph{pop stack} is a container in which elements can be stacked on top of each other, on which two operations are defined: the \texttt{PUSH} operation, which inserts an element into the stack, and the \texttt{POP} operation, which extracts \emph{all the elements} from the stack. The difference between a pop stack and a (classical) stack lies therefore in the way the elements are removed from the stack.

When using a stack or a pop stack to sort a permutation, the elements of the permutations are usually processed from left to right, and either the current element of the permutation is pushed into the stack, or the topmost element (or all the content) of the stack is popped into the output, again from left to right. We now introduce a new kind of pop stack, by allowing one more operation: this is the \texttt{BYPASS} operation, which takes the current element of the input permutation and moves it directly into the output, in the next available position. More formally, given a permutation $\pi=\pi_1 \pi_2 \dots \pi_n$, a pop stack with bypass has the following allowed operations:
\begin{itemize}
	\item[\texttt{PUSH:}] insert the current element of the input into the pop stack, on top of all the other elements (if there are any);
	\item[\texttt{POP}:] remove all the elements in the pop stack, from top to bottom, sending them into the output;
	\item[\texttt{BYPASS}:] output the current element of the input.    
\end{itemize}

Our goal is to use a pop stack with bypass in order to sort permutations. It is not too difficult to realize that not every permutation can be sorted.

\begin{prop}
Let $\pi$ be a permutation such that $231\le\pi$ or $4213\le\pi$. Then $\pi$ cannot be sorted using a pop stack with bypass.
\end{prop}
\begin{proof}
Suppose that $\pi$ contains the pattern $231$, and let $bca$ be an occurrence of $231$ in $\pi$. We analyze all possible configurations that can occur during the sorting process.
In order to obtain the identity permutation, neither $b$ nor $c$ can reach the output before $a$, therefore they must enter the pop stack and wait until $a$ is the current element of the input. However, this implies that $b$ and $c$ are inserted into the pop stack in this order, so they will necessarily appear in the output in reverse order, which means that the output will be unsorted. 

Now suppose that $\pi$ contains the pattern $4213$, and let $dbac$ be an occurrence of $4213$ in $\pi$. Using a similar argument as before, we observe that, in order to sort the output, both $d$ and $b$ must enter the pop stack and wait for $a$ to be the current element of the input. Since the \texttt{POP} operation removes all the elements currently in the pop stack, $b$ and $d$ will reach the output before or after $c$ (that is, the output will contain either the subsequence $bdc$ or the subsequence $cbd$). This implies that the output will not be the identity permutation.
\end{proof}

We will now prove the converse of the above proposition, by introducing an optimal sorting algorithm, that we call \textsf{PSB} (an acronym for ``PopStacksort with Bypass"), by which we can sort all permutations in the class $\Av(231,4213)$.

\begin{algorithm}
	$S:=\emptyset$\;
	$i:=1$\;
	\While{$i\leq n$}
	{
		\If{$S=\emptyset$ or $\pi_i =\textnormal{\texttt{TOP}}(S)-1$}
		{
			\texttt{PUSH}\;
		}
		\ElseIf{$\pi_i <\textnormal{\texttt{TOP}}(S)-1$}
		{
			\texttt{BYPASS}\;
		}
		\Else
		{
			\texttt{POP}\;
			\texttt{PUSH}\;
		}
		$i:=i+1$\;		
	}
	\texttt{POP}\;
	\caption{\textsf{PSB} ($S$ is the pop stack; \texttt{TOP}(S) is the current top element of the pop stack; 
		$\pi=\pi_1 \cdots \pi_n$ is the input permutation).}\label{queuesort}
\end{algorithm}

\begin{figure}[h]
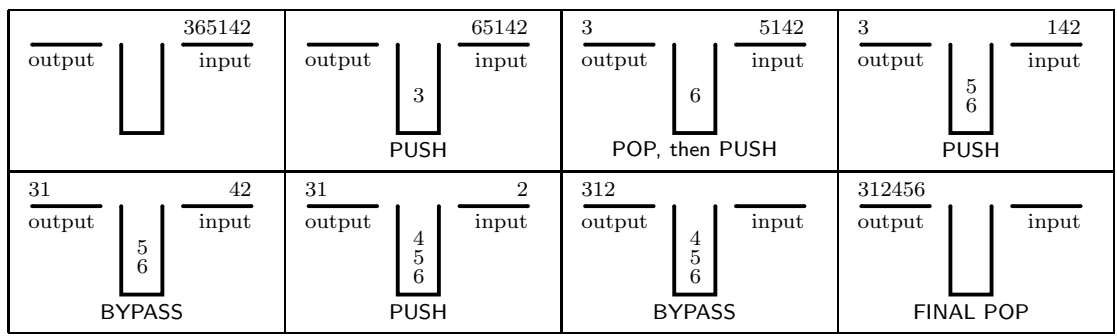

	\centering
	
\begin{tabular}{|c|c|c|c|}
\hline
\begin{onestack}
\fillstack{}{}{365142}
\end{onestack}
&
\begin{onestack}
\fillstack{}{3}{65142}
\op{\textsf{PUSH}}
\end{onestack}
&
\begin{onestack}
\fillstack{3}{6}{5142}
\op{\textsf{POP, then PUSH}}
\end{onestack}
&
\begin{onestack}
\fillstack{3}{65}{142}
\op{\textsf{PUSH}}
\end{onestack}
\\
\hline
\begin{onestack}
\fillstack{31}{65}{42}
\op{\textsf{BYPASS}}
\end{onestack}
&
\begin{onestack}
\fillstack{31}{654}{2}
\op{\textsf{PUSH}}
\end{onestack}
&
\begin{onestack}
\fillstack{312}{654}{}
\op{\textsf{BYPASS}}
\end{onestack}
&
\begin{onestack}
\fillstack{312456}{}{}
\op{\textsf{FINAL POP}}
\end{onestack}
\\\hline
\end{tabular}
    
	\caption{Using \textsf{PSB} to sort $365142$}
	\label{example_PSB}

\end{figure}

The algorithm \textsf{PSB} maintains elements in the pop stack only when they are consecutive in value (and increasing from top to bottom), which ensures at least that a \texttt{POP} operation is not directly responsible for the possible failure of the sorting procedure. 
An example of how $\textsf{PSB}$ works is shown in Figure \ref{example_PSB}, where the input permutation (on the right of the pop stack) is not sortable.

To prove that \textsf{PSB} actually sorts every permutation in $\Av(231,4213)$ we need some preliminary lemmas. In what follows, we will denote by $\psb(\pi)$ the output of \textsf{PSB} on the permutation $\pi$. 

\begin{lemma}
\label{noInv}
Let $a,b$ be two positive integers such that $a<b$, and let $\pi$ be a permutation. If the pair $(a,b)$ is a noninversion in $\pi$, then it remains a noninversion in $\psb(\pi)$ (that is, \textsf{PSB} does not create new inversions).
\end{lemma}
\begin{proof}
Let $a<b$, and suppose that $a$ appears to the left of $b$ in $\pi$. Consider the configuration in which $a$ is the current element of the input, and suppose that $a$ enters the pop stack (otherwise there is nothing to prove). If $a$ is still in the pop stack when $b$ is processed, then $b$ is certainly larger than the top of the pop stack, and so a \texttt{POP} operation is performed, which is enough to ensure that $a$ is also to the left of $b$ in $\psb(\pi)$.  
\end{proof}

We also propose a (rather technical) generalization of the previous lemma, which is not actually needed in the proof of the next Proposition \ref{other_direction}, but which instead will be extremely useful in Section \ref{classes}.

\begin{lemma}\label{beta}
	Given a permutation $\pi$, suppose that $\psb(\pi )$ contains the subsequence $m\beta$, where $m\in \mathbf{N}$ is larger than each element of the sequence $\beta$. Then $m\beta$ is also a subsequence of $\pi$. 
\end{lemma}
\begin{proof}
	The lemma is obvious when $\beta$ is empty, so suppose that $\beta$ contains at least one element. Thanks to the previous lemma, $\pi$ contains the subsequence $m\tilde{\beta}$, for some rearrangement of the elements of $\beta$. Since $m$ reaches the output before the elements of $\beta$, there must be an element $M>m$ before $\tilde{\beta}$ in $\pi$. As a consequence, we observe that, when an element of $\tilde{\beta}$ is the current element of the input, the top of the pop stack is certainly strictly larger than $m$ (which is in turn larger than all elements of $\tilde{\beta}$), and therefore the output will also contain the sequence $\tilde{\beta}$. This means that actually $\tilde{\beta}=\beta$.
\end{proof}

\begin{lemma}
\label{LTRinStack}
Suppose that $a$ is an element of a permutation $\pi$ that enters the pop stack during the execution of \textsf{PSB}. Then there exists a (necessarily unique) nonnegative integer $k\geq 0$ such that $\pi$ contains the decreasing subsequence of (not necessarily adjacent) consecutive elements $a+k, a+k-1$, \dots, $a$ and $a+k$ is the rightmost left-to-right maximum of $\pi$ (weakly) to the left of $a$.
In particular, when $k=0$, we see that the left-to-right maxima of $\pi$ are precisely the elements that are at the bottom of the pop stack at some point during the sorting procedure.
\end{lemma}
\begin{proof}
We argue by induction on the position of the element $a$ in $\pi$.

If $a$ is the first element of $\pi$, then $a$ is a left-to-right maximum and enters the pop stack.

Now suppose that $a$ is the $i$-th element of $\pi$ and that $a$ enters the pop stack. This may happen in two distinct ways. The first case is when $a$ is larger than the current top of the pop stack (and so it induces a \texttt{POP} operation). In such a case, we see that $a$ is larger than \emph{every} element inside the pop stack (since the elements in the pop stack are consecutive), and so in particular it is larger than the current element lying at the bottom of the pop stack, which (by the induction hypothesis) is a left-to-right maximum of $\pi$. Therefore, $a$ is a left-to-right maximum of $\pi$. The second case occurs when $a$ is pushed into the pop stack on top of a nonempty sequence of elements. In such a case, the elements in the pop stack (from top to bottom) are $a+1,a+2,\ldots ,a+k$ (for some $k\geq 1$). This means that $\pi$ contains the subsequence $a+k,\ldots ,a+1,a$, and that $a+k$ is a left-to-right maximum (by the induction hypothesis). Moreover, there cannot be elements larger than $a+k$ between $a+k$ and $a$ in $\pi$, otherwise any such element would have induced a \texttt{POP} operation, and so $a+k$ could not be inside the pop stack when $a$ is processed. This means that $a+k$ is the rightmost left-to-right maximum to the left of $a$.
\end{proof}

\begin{prop}\label{other_direction}
Let $\pi \in \Av(231,4213)$. Then $\psb(\pi)$ is an identity permutation.
\end{prop}
\begin{proof}
Let $\pi$ be a permutation such that $\psb(\pi)$ is not an identity permutation. This means that there exist two elements $a$, $b$ such that $a<b$ and where $b$ appears to the left of $a$ in $\psb(\pi)$. We want to prove that $\pi$ contains $231$ or $4213$.

By \cref{noInv}, the element $b$ must occur to the left of $a$ in $\pi$ as well. We now focus on the configuration when $b$ is sent to the output. This may happen by means of either a \texttt{POP} operation or a \texttt{BYPASS} operation.

In the former case, the element $b$ is sent to the output from the pop stack; therefore, the current element of the input, say $c$, must be larger than \texttt{TOP}$(S)$. Since the content of the pop stack consists of consecutive elements by construction, then $b<c$, and the elements $b$, $c$, $a$ form an occurrence of the pattern $231$ in $\pi$.

In the latter case, let $c=$\texttt{TOP}$(S)$. This implies $b<c-1$. The element $c-1$ must appear somewhere in $\pi$. If it appears to the left of $c$, then $(c-1)cb$ forms an occurrence of the pattern $231$ in $\pi$. Clearly $c-1$ cannot occur between $c$ and $b$, otherwise it would be in the pop stack by \cref{LTRinStack}. If $c-1$ is between $b$ and $a$, then $b(c-1)a$ forms an occurrence of the pattern $231$ in $\pi$. Finally, if $c-1$ appears to the right of $a$ in $\pi$, then $cba(c-1)$ forms an occurrence of the pattern $4213$ in $\pi$.

Therefore, if $\pi\in \Av(231,4213)$, then $\psb(\pi)$ is an identity permutation.
\end{proof}

\begin{figure}[h]
	\centering
	\includegraphics[scale=0.6]{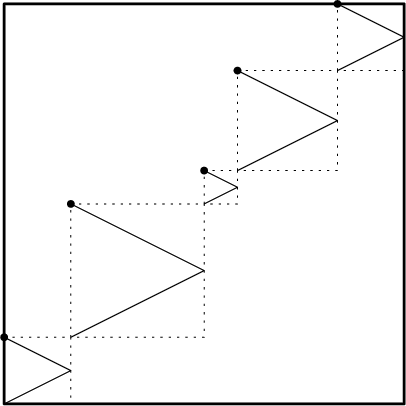}
	\caption{The bullets indicate that each summand begins with its maximum.}
	\label{av_231_4213}
\end{figure}

The class of permutations sortable using a pop stack with a bypass is, of course, a superclass of those sortable using a classical pop stack, which is $\Av (231,312)$, as shown in \cite{AN}. From a geometric point of view, the structure of permutations in the class $\Av(231,4213)$ is depicted in Figure \ref{av_231_4213}. More formally, each such permutation can be expressed in a unique way as the direct sum of permutations avoiding 231 and 213 starting with their maximum. Moreover, in each direct summand, the entries of the decreasing sequence are those entering the pop stack, whereas the entries of the increasing sequence are those that bypass the pop stack.

\section{A bijective link with a class of restricted Motzkin paths}\label{motzkin_link}

The coefficients $(|\Av_n (231,4213)|)_n$ are the odd-indexed Fibonacci numbers, that is, the sequence A001519 in \cite{Sl}, as shown in \cite{A}. For our purposes, Fibonacci numbers are precisely sequence A000045 in \cite{Sl}, and so they are defined by the usual recurrence $F_{n}=F_{n-1}+F_{n-2}$ (for $n\geq 2$), with initial conditions $F_0=0$ and $F_1=1$.

It is possible to give a bijective proof of this enumerative result by providing a link with a class of restricted Motzkin paths, whose enumeration can be easily carried out using standard techniques.

\bigskip

To be more precise, there is a very nice bijection between the permutations sortable by \textsf{PSB} and a subset of ternary words that are counted by the odd-indexed Fibonacci numbers.  The bijection is to take the operation(s) that each entry of each sortable permutation utilizes in \textsf{PSB} and convert it to a letter of the word.  

For each permutation $\pi$, the \emph{sorting word} of $\pi$ will be defined entry-wise using the operation of \textsf{PSB} used when the entry is the next in the input to be acted on as follows:
\begin{itemize}
\item[]\texttt{PUSH}:  $0$
\item[]\texttt{BYPASS}: $1$
\item[]\texttt{POP}, then \texttt{PUSH}: $2$
\end{itemize}    
Note that since each entry of the permutation will either be pushed into the pop stack or bypass the pop stack exactly once, the sorting word will have the same size as the sortable permutation.

\begin{figure}[t]
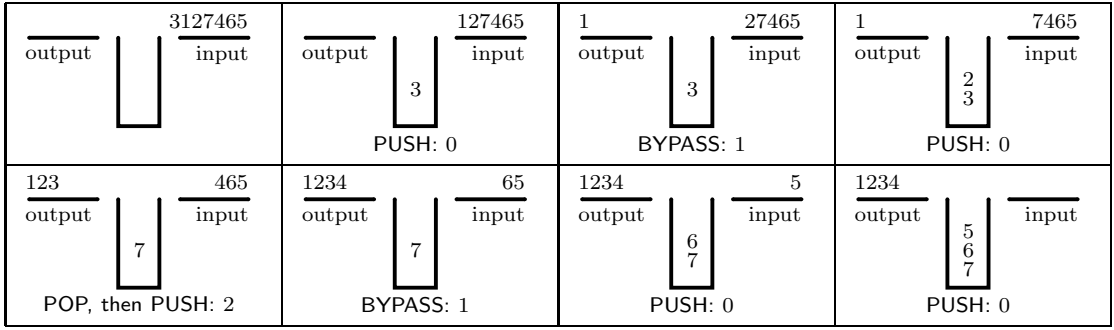

\begin{center}

\begin{tabular}{|c|c|c|c|}
\hline
\begin{onestack}
\fillstack{}{}{3127465}
\end{onestack}
&
\begin{onestack}
\fillstack{}{3}{127465}
\op{\textsf{PUSH:} $0$}
\end{onestack}
&
\begin{onestack}
\fillstack{1}{3}{27465}
\op{\textsf{BYPASS:} $1$}
\end{onestack}
&
\begin{onestack}
\fillstack{1}{32}{7465}
\op{\textsf{PUSH:} $0$}
\end{onestack}
\\
\hline
\begin{onestack}
\fillstack{123}{7}{465}
\op{\textsf{POP, then PUSH:} $2$}
\end{onestack}
&
\begin{onestack}
\fillstack{1234}{7}{65}
\op{\textsf{BYPASS:} $1$}
\end{onestack}
&
\begin{onestack}
\fillstack{1234}{76}{5}
\op{\textsf{PUSH:} $0$}
\end{onestack}
&
\begin{onestack}
\fillstack{1234}{765}{}
\op{\textsf{PUSH:} $0$}
\end{onestack}
\\\hline
\end{tabular}

\caption{Sorting $3127465$ using \textsf{PSB}}
\label{fig_3127465}
\end{center}
\end{figure}

For example, the permutation $\pi= 3127465$ has sorting word $0102100$ as shown in Figure~\ref{fig_3127465}.

Let $W$ be the set of ternary words on $\{0,1,2\}$ such that each word: 
    \begin{enumerate}
        \item begins with $0$,
        \item ends with $0$ or $2$, and
        \item avoid the consecutive pattern $12$.
    \end{enumerate} 

\begin{prop}~\label{P:sorting_words}
    The sorting words of permutations sortable by \textsf{PSB} are exactly the ternary words of $W$. 
\end{prop}

\begin{proof}
    To see that these conditions are necessary is a straightforward application of the algorithm \textsf{PSB} which prioritizes pushing entries into the pop stack over bypassing when appropriate for sorting.  The first entry of the permutation must be pushed into the pop stack, so every sorting word begins with $0$.  The last entry cannot bypass the stack as if this would allow for sorting, the last entry must be one less than the current top entry of the pop stack or the pop stack must be empty, so the sorting word cannot end with a $1$.  Finally, the consecutive pattern $12$ would indicate a bypass followed by a pop, which again means that the entry that bypassed the stack could have been pushed to the top of the pop stack and then popped with the rest of those entries for the same output. This proves that the set of sorting words of sortable permutations is contained in $W$.
    
   	Conversely, given $w\in W$, we now show that $w$ is the sorting word of a sortable permutation by explicitly constructing the unique sortable permutation $\pi$ having $w$ as sorting word. To do so, we start by observing that $w$ can be factorized into subwords as $w=w_1 w_2 \cdots w_k$ such that, for all $i\geq 2$, each $w_i$ starts with 2 and (if it has size strictly greater than 1) the remaining letters are 0's and 1's, with a 0 at the end; the factor $w_1$ has exactly the same form, except that it starts with 0. The form of this factorization depends on the defining properties of $W$, in particular on the fact that there are no consecutive occurrences of 12. Denote with $\ell_i$ the size of $w_i$, for each $i$.

    The permutation $\pi$ that we are going to construct can be factorized in terms of its left-to-right maxima as $\pi =\alpha_1 \alpha_2 \cdots \alpha_k$ where, for each $i$, $\alpha_i$ starts with a left-to-right maximum. 
    Since every left-to-right maximum (except for the first one) pops the stack, the output of \textsf{PSB} on $\pi$ is $\mathbf{PSB}(\pi )=\mathbf{PSB}(\alpha_1 )\mathbf{PSB}(\alpha_2 )\cdots \mathbf{PSB}(\alpha_k )$.
    In view of Lemma \ref{LTRinStack}, the set containing the initial letter 0 and each letter 2 in a sorting word correspond to the left-to-right maxima of the permutation associated with that sorting word.
    Thus for $w$ to be the sorting word of $\pi$, then $\alpha_i$ need to have size $\ell_i$, for each $i$. Moreover, since we require $\pi$ to be sortable, each $\alpha_i$ must contain all the integers from $\ell_{i-1}+1$ to $\ell_i$.

    The above discussion allows us to finally describe a procedure to construct $\pi$ starting from $w$. We successively replace the letters of $w$ with positive integers, starting from 1 and increasingly, as follows. For $i$ running from 1 to $k$, scan $w_i$ from left to right and replace each 1 with an integer (increasingly from left to right); then replace with integers (increasingly from right to left) the remaining letters of $w_i$. As a consequence of the above discussion, the resulting permutation $\pi$ is sortable and has $w$ as its sorting word.       
    
    For instance, starting from $w=\underbrace{0110}_{w_1}\underbrace{210}_{w_2}\underbrace{2}_{w_3}\underbrace{2010}_{w_4}\underbrace{2}_{w_5}\in W$, one obtains the permutation $\pi =\underbrace{4\ 1\ 2\ 3}_{\alpha_1}\ \underbrace{7\ 5\ 6}_{\alpha_2}\ \underbrace{8}_{\alpha_3}\ \underbrace{12\ 11\ 9\ 10}_{\alpha_4}\ \underbrace{13}_{\alpha_5}$.
    \end{proof}
    
The permutations sortable by \textsf{PSB} can similarly be placed in bijection with a restricted class of Motzkin paths.  Recall that a \emph{Motzkin path} is a lattice path from $(0,0)$ to $(0,n)$ that never goes below the $x$ -axis which takes steps of the following types:
\begin{itemize}
    \item Up (NE): $(1,1)$
    \item Down (SE): $(1,-1)$
    \item Horizontal (E): $(1,0)$
\end{itemize}

For each permutation $\pi$, the \emph{sorting path} of $\pi$ will be defined entry-wise.  Specifically, for each entry $\pi_i$, use the operation of \textsf{PSB} utilized when $\pi_i$ is the next in the input to be acted on as follows:
\begin{itemize}
    \item[]\texttt{PUSH}:  Up step
    \item[]\texttt{BYPASS}: Horizontal step
    \item[]\texttt{POP}, then \texttt{PUSH}: Down steps until reaching the $x$-axis, followed by an Up Step
    \item[]Completion: Down steps until the $x$-axis is reached.
\end{itemize}  
Note that a Motzkin path corresponding to a sorting word of size $n$ will have exactly $n$ total Up and Horizontal steps combined.

\begin{figure}[h!]
    \centering
    \begin{tikzpicture}[scale=0.75]
    \draw [help lines] (0,0) grid (12,4);
    \draw [blue, line width = 1pt ] (0,0) -- (1,1) --(2,1) --(3,2) --(4,1) -- (5,0)
        --(6,1) -- (7,1) --(8,2) -- (9,3) --(10,2) --(11,1) --(12,0);
    \end{tikzpicture}
    \caption{The Motzkin path $UHUDDUHUUDDD$ corresponding to $\pi= 3127465$}
    \label{fig:M_path}
\end{figure}
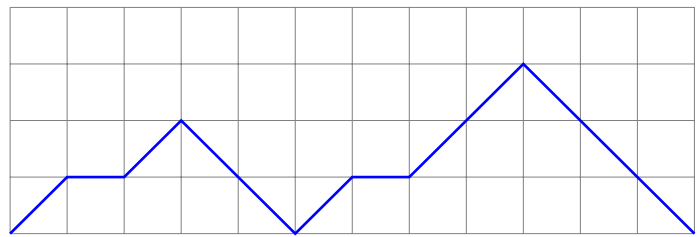

Returning to our previous example where $\pi= 3127465$, we can see that $\pi$ has sorting path $UHUDDUHUUDDD$ shown in Figure~\ref{fig:M_path} based on its sorting sequence from \textsf{PSB} as was shown in Figure~\ref{fig_3127465}.  

Using the same methods as in Proposition~\ref{P:sorting_words}, the permutations sortable by \textsf{PSB} can be shown to be in bijection with the set of restricted Motzkin paths $M$ defined below.

Let $M$ be the subset of Motzkin paths where each path $p \in M$ satisfies the following added restrictions.
    \begin{enumerate}
        \item $p$ begins with an Up step,
        \item $p$ ends with a Down step,
        \item $p$ never has a Horizontal step immediately preceded or followed by a Down step,
        \item when $p$ has a Down step, that step is immediately followed by as many Down steps as necessary to reach the $x$-axis.
    \end{enumerate}

\begin{prop}~\label{P:sorting_paths}
    The sorting paths of permutations sortable by \textsf{PSB} are exactly the Motzkin paths of $M$. 
\end{prop}

For enumeration purposes, recalling that $\Av_n (231,4213)$ is the set of permutations sortable by \textsf{PSB} of size $n$, denote by 
\begin{itemize}
    \item[$W_n$] the corresponding set of sorting words of $W$ size $n$, and 
    \item[$M_n$] the corresponding set of sorting paths of $M$ with exactly $n$ total Up and Horizontal steps combined.
\end{itemize}

The enumeration of the class of permutations $\Av(231, 4213)$ was originally given by Atkinson~\cite{A}, but given the new context that defines these same permutations as those sortable by \textsf{PSB}, we can use either the bijection from Proposition~\ref{P:sorting_words} or Proposition~\ref{P:sorting_paths} to reconstruct this result using recursion.

\begin{theorem}~\label{T:enumeration} (Atkinson~\cite{A})
    The number of permutations sortable by \textsf{PSB} of size $n$ is given by the odd indexed Fibonacci number $F_{2n-1}$ where we define $F_{-1}=1$ for consistency.
\end{theorem}

\begin{proof}
    The only Motzkin path with exactly $n=0$ total Up and Horizontal steps is the empty path, so $|M_0| = 1= F_{-1}$.  Furthermore, the only path in $M_1$ is $UD$, so $|M_1| = 1 = F_{1}$. For $n \geq 2$, each Motzkin path in $M_n$ can begin one of three ways $UD$, $UpD$, $UHH\cdots HqD$ where $p,q$ begin with a $U$ step and the last $D$ shown in each prefix is the first return of the corresponding path to the $x$-axis, i.e. the $D$ shown corresponds to the pop of the entry corresponding to the first $U$.  Thus, $p,q$ are themselves paths in $M$.  
    
    Removing the $UD$ in the first case creates a bijective map between these paths of $M_n$ and all paths of $M_{n-1}$.  Similarly, if a path begins $UpD$ where $p$ begins with a $U$ step and the last $D$ shown in each prefix is the first return of the corresponding path to the $x$-axis, removing the first $U$ and the last $D$ of this prefix gives a bijection between these paths of $M_n$ and all paths of $M_{n-1}$.  Finally, in the case of the paths that begin $UHH\cdots HqD$, if we remove the first $U$, the initial sequence of consecutive $H$ steps and the last $D$ of this prefix (so that the new prefix is $q$), these paths are in bijection with the Motzkin paths of $\bigcup_{i=1}^{n-2} M_{i}$ as $q$ is nonempty.

    By induction and some Fibonacci identities,
    \begin{align*}
        |M_n| &= 2|M_{n-1}| + \sum_{i=1}^{n-2}|M_i| \\
        &= F_{2(n-1)-1} + \sum_{i=1}^{n-1}F_{2i-1} \\
        &= F_{2n-3}+ F_{2n-2} \\
        &= F_{2n-1}.
    \end{align*}
\end{proof}

Alternatively, we can see that the sorting words have the same recursive behavior or look at the regular expression for the set of sorting words.  We give a proof sketch using the latter here.
\begin{proof}
The set of sorting words for sortable permutations is a regular language. In fact, it is generated by the regular expression $0(0+2+1^+ 0)^*$ (where, in particular, the notation $1^+$ is a shortcut for $11^*$). A standard application of the classical Sch\"utzenberger methodology~\cite{CS} gives immediately the generating function $F(x)$ of the sequence that counts the sorting words of sortable permutations with respect to the length, which is
\[
    F(x)=x\cdot \frac{1}{1-\left( 2x+x\cdot \frac{x}{1-x} \right)}=\frac{x(1-x)}{1-3x+x^2} .
\]

It is then easy to see that $F(x)$ is the generating function of odd-indexed Fibonacci numbers (with the correct initial conditions). 
\end{proof}

\section{Preimages under \textsf{PSB}}\label{preimages_perms}

In this section we investigate the preimages of a generic permutation $\sigma$ under \textsf{PSB}. More precisely, since the algorithm \textsf{PSB} induces a map $\psb$ from the set of all permutations to itself, we may ask to determine the set of permutations $\psb^{-1}(\sigma)$ of all permutations whose output under \textsf{PSB} is $\sigma$. This kind of investigation is somehow classical for \texttt{Stacksort} \cite{B-M,D1}, and has been recently carried out also for \texttt{Queuesort} \cite{CF1} and \texttt{Bubblesort} \cite{BCF}.

We start by providing an alternative description of \textsf{PSB} that focuses on the left-to-right maxima of the input permutation. To this aim, we will use the notion of shuffle, which we define next.

Let $\rho$, $\sigma$ be two sequences of distinct integers. We say that a sequence $\tau$ is a \emph{shuffle} of $\rho$ and $\sigma$ if $\tau$ contains both $\rho$ and $\sigma$ as subsequences and contains no elements other than those of $\rho$ and $\sigma$.
We denote the set of all possible shuffles of $\rho$ and $\sigma$ with $\rho\shuffle\sigma$. 
As an example, if $\rho =31$ and $\sigma =24$, then the set of all shuffles of $\rho$ and $\sigma$ is $\rho \shuffle \sigma =\{ 3124,3214,3241,2314,2341,2431\}$. 

\bigskip

We can now give an alternative description of \textsf{PSB}.

Let $\pi\in S_n$, and let $m_1,\dots ,m_k$ be its left-to-right maxima. For every $i=1, \dots, k$, suppose that $m_i (m_i-1)\cdots (m_i-j_i)$ is the longest decreasing subsequence of consecutive elements starting with $m_i$ and entirely to the left of $m_{i+1}$ (if it exists). Then we can write $\pi$ as $\pi =m_1 A_1 \cdots m_k A_k$, with $A_i \in ( (m_i-1)\cdots (m_i - j_i))\shuffle P_i$, where the $P_i$'s collect all the remaining elements of $\pi$.
Using \cref{LTRinStack} and the description of \textsf{PSB}, we obtain
$\psb(\pi )=P_1 (m_1 - j_1)\cdots m_1 P_2 (m_2-j_2)\cdots m_2 \cdots P_k (m_k - j_k) \cdots m_k$.

For example, for the permutation $\pi = 635247198$, we see that $m_1=6$, $m_2=7$, $m_3=9$, $j_1=2$, $j_2=0$, and $j_3=1$. Moreover, $P_1=32$ and $P_2=1$. Furthermore, $P_3$ is empty, hence we obtain $\psb(\pi)=324561789$. Notice that $3524$ is one of the possible shuffles of $54$ and $32$, but there exist others, such as $3542$. We remark that any shuffle of $54$ and $32$ (in the input permutation $\pi$) will give the same output. For instance, setting $\sigma=635427198$, we get $\psb(\sigma)=324561789=\psb(\pi)$.

The description above allows us to understand the relation between the left-to-right maxima of a permutation and those of its preimages.

\begin{prop}
Let $\pi$, $\sigma$ be permutations such that $\psb(\pi )=\sigma$. If $m$ is a left-to-right maximum of $\pi$, then $m$ is a left-to-right maximum of $\sigma$ as well.
\end{prop}
\begin{proof}
The proof is an immediate consequence of the description of \textsf{PSB} given above.
In fact, setting $\pi=m_1 A_1\cdots m_k A_k$ with the same notation as above, we see that, for every $i=1,\dots ,k$, the element $m_i$ is greater than all the elements of $A_i$. Since $\sigma=\psb (\pi )=P_1 (m_1 - j_1)\cdots m_1 P_2 (m_2-j_2)\cdots m_2 \cdots P_k (m_k - j_k) \cdots m_k$, then $m_i$ is clearly a left-to-right maximum also of $\sigma$.
\end{proof}

Since sequences of consecutive left-to-right maxima play an important role in the algorithm, we provide a description of the output permutation that conveniently highlights such sequences. For a given permutation $\sigma$, we write $\sigma=\mu_1 B_1\cdots \mu_{k-1} B_{k-1} \mu_k$, where the $\mu_i$'s are the maximal sequences of consecutive left-to-right maxima of $\sigma$, and the $B_i$'s (possibly empty) collect all remaining elements. For each $i$, we set $\mu_i=m_{i,1}\cdots m_{i,\ell_i}$.

Since \textsf{PSB} always outputs a permutation ending with its maximum, we will be interested exclusively in such permutations.

We now introduce a recursive algorithm that generates all preimages of a given permutation $\sigma$. We start by giving an informal description of the algorithm, in order to better understand how it works. Notice that our algorithm is actually defined on sequences of distinct integers (not necessarily permutations): this allows us to recursively execute it on subsequences of elements of a permutation.

Suppose that $\sigma =\sigma_1 \sigma_2 \cdots \sigma_n =\alpha \mu_k$, where $\mu_k$ is the maximum suffix of consecutive left-to-right maxima of $\sigma$ (and $\alpha$ is the remaining prefix). For each entry $m$ in $\mu_k$, construct permutations as follows. First, remove the suffix of left-to-right maxima starting with $m$. Then reinsert the removed elements into the (remaining prefix of the) permutation in all possible ways, according to the following rules:
\begin{itemize}
	\item the elements are reinserted in decreasing order;
	\item the maximum (i.e. $n$) is inserted to the immediate right of one of the remaining left-to-right maxima of $\sigma$ (i.e. a left-to-right maximum to the left of $m$ in $\sigma$)
	or at the beginning of $\sigma$;
	\item the minimum (i.e. $m$) is inserted somewhere to the right of $m-1$. 
\end{itemize}

\begin{figure}[htb]
    \begin{minipage}[t]{.1\textwidth}
    \phantom{a}
    \end{minipage}
    \begin{minipage}[t]{.2\textwidth}
        \centering
        \includegraphics[width=\textwidth]{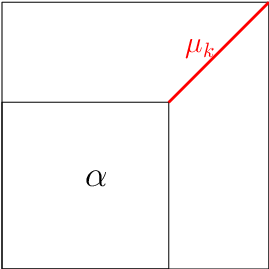}
    \end{minipage}
    \hfill
    \begin{minipage}[t]{.2\textwidth}
        \centering
        \vspace{-2cm}
        \includegraphics[scale=0.4]{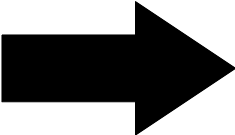}
    \end{minipage}
    \hfill
    \begin{minipage}[t]{.2\textwidth}
        \centering
        \includegraphics[width=\textwidth]{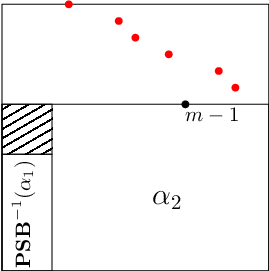}
    \end{minipage}
    \begin{minipage}[t]{.1\textwidth}
    \phantom{a}
    \end{minipage}
    \caption{Computing one preimage of $\sigma =\alpha \mu_k$ when $m$ is the first (i.e. smallest) element of $\mu_k$.}
    \label{fig:preimages}
\end{figure}

At this point, consider the prefix of all elements strictly before $n$ and (recursively) compute all its possible preimages. Figure \ref{fig:preimages} illustrates how the algorithm works for a specific choice of $m$. In the figure, the permutation $\sigma =\alpha \mu_k$ is decomposed as $\sigma =\alpha_1 \alpha_2 \mu_k$, where $\alpha_1$ is the prefix of $\alpha$ ending with the left-to-right maximum to the right of which $n$ is reinserted.

We illustrate the above algorithm by computing the preimages of the permutation 3154267. In this case the maximum suffix of consecutive left-to-right maxima is 67. So we have two families of preimages.

\begin{itemize}
    \item If we remove 6 and 7, we have three possible positions where to reinsert 7 (at the beginning, right after 3 or right after 5) and, for each of them, three possible positions where to reinsert 6 (right after 5, 4 or 2). If we reinsert 7 at the beginning, we thus get the three preimages 7315642,7315462 and 7315426. If we reinsert 7 after 3, we get the three preimages 3715642, 3715462 and 3715426. If we reinsert 7 after 5, after having computed the unique preimage of 315, which is 351, we get the three preimages 3517642, 3517462 and 3517426.
    \item If we remove only 7, we need to (recursively) compute all the preimages of the prefix 315426, which turns out to be just one, and append 7. In this way, we obtain the further preimage 3516427.
\end{itemize}

\begin{prop}\label{preimages}
	The above algorithm computes all the preimages of a given permutation.
\end{prop}

\begin{proof} \emph{(Skecth.)}\quad Let $\sigma \in S_n$. It is easy to realize that all the permutations produced by the above described algorithm are indeed preimages of $\sigma$ under $\mathsf{PSB}$.

Conversely, let $\pi \in \psb^{-1}(\sigma )$. If we decompose $\pi$ as $\pi =LnR$, we observe that the elements in $L$ are processed by $\mathsf{PSB}$ independently from the rest of $\pi$, hence $\psb (L)$ is a prefix of $\sigma$. When $n$ is pushed into the pop stack (clearly at the bottom of it), each of the remaining elements of $\pi$ (i.e., the elements of $R$) is either pushed into the pop stack or bypasses it. After all elements of $\pi$ are processed, the elements in the pop stack are poured into the output. From the definition of $\mathsf{PSB}$ it then follows that the elements appearing in the pop stack after $n$ is processed constitute the maximal decreasing subsequence of $\pi$ made of consecutive elements starting with $n$ (call it $\rho$). Notice that the elements of $\rho$ need not be adjacent in $\pi$.  Notice that all the remaining elements of $R$ appear in $\sigma$ in the same order as they appear in $\pi$. In particular, as a consequence of the maximality of $\rho$, the maximum of the elements of $R$ bypassing the pop stack necessarily appears to the left of the minimum of $\rho$ in $R$. Collecting all the above information tells us that $\pi$ must have precisely the form prescribed by our algorithm for preimages.     
\end{proof}

\bigskip

In the rest of the present section we focus on the sets of permutations having a fixed number of preimages through \textsf{PSB}.

Given $n\ge 0$, let $C_n^{(k)}=\{\sigma \in S_n\mid |\psb^{-1}(\sigma )| = k\}$, and $c_n^{(k)}=|C_n^{(k)}|$.

\begin{prop}
\[
C_n^{(0)}=\{\sigma =\sigma_1 \cdots \sigma_n \mid \sigma_n \neq n\}
\]
and 
\[
c_n^{(0)}=(n-1) (n-1)!.
\]
\end{prop}
\begin{proof}
We have already observed that any permutation in the image of $\psb$ ends with its maximum. On the other hand, if a permutation $\sigma$ ends with its maximum $n$, then by Proposition \ref{preimages} it has at least one preimage under $\psb$. For instance, using the same notation as in the above algorithm to compute preimages, if $\sigma =\alpha \mu_k$, with $\mu_k =\mu_{k,1}\mu_{k,2}\cdots \mu_{k,\ell_k}=\mu_{k,1}\beta \neq \emptyset$ and $\beta \neq \emptyset$, then $\rho =\beta^r \alpha \mu_{k,1}$ is a preimage of $\sigma$ (the case $\beta =\emptyset$ requires a little bit more care, but can still be solved using our algorithm).
Therefore, a permutation has no preimages if and only if it does not end with its maximum.

The number of permutations with no preimages is thus $(n-1)(n-1)!$, since to construct such a permutation we have to choose its final element among the $n-1$ possible values, while the other elements can be arranged in all possible ways.
\end{proof}

\begin{prop}
For $n\geq 3$, the set $C_n^{(1)}$ consists of all permutations of size $n$ ending with $n$ whose left-to-right maxima are consecutive and nonadjacent. More formally,  
\begin{gather*}\label{preim1}
C_n^{(1)}=\{\pi=\pi_1\cdots\pi_n\mid \pi_n=n \text{ , there exists $k\ge 0$ such that } LTR(\pi)=\{n-k,\dots,n\}\\
\text{ and for every $\pi_i$, $\pi_j \in LTR(\pi)$, $\pi_i\neq\pi_j$, we have } |j-i|>1 \} .
\end{gather*}

Moreover
\[
c_n^{(1)} = \sum_{k=2}^{\lceil\frac{n}{2}\rceil} (n-k)! \binom{n-k-1}{k-2}.
\]
\end{prop}
\begin{proof}
Let $\pi=\pi_1\cdots\pi_n$ be a permutation such that $\pi_n=n$ and the set of its left-to-right maxima is a subsequence of consecutive and nonadjacent entries of $\pi$.
Then, using our notation, we have $\pi=(n-k) B_k (n-k+1) B_{k-1}\cdots (n-1) B_{1} n$ (modulo a slight change in the indexing), and we know by \cref{preimages} that the only preimage of $\pi$ is $(n-k) (n-k+1) B_k (n-k+2) B_{k-1}\cdots n B_{1}$.

On the other hand, we now show that any permutation which has not the form described in the statement of the proposition cannot have exactly one preimage. Clearly, if $\pi$ does not end with $n$ then it has no preimages. So let $\pi=\pi_1\cdots \pi_n$ be a permutation with $\pi_n = n$ and such that $LTR(\pi)$ is not an interval. Let $l, r\in LTR(\pi)$ such that $l+1, l+2,\dots, r-1\notin LTR(\pi)$ and $r+1, r+2,\dots, n-1, n\in LTR(\pi)$ (in other words, $l$ and $r$ identify the rightmost gap in $LTR(\pi)$). Thus we can express $\pi$ as $\pi =\pi_1 L l M r R_1 (r+1) R_2 (r+2)\cdots R_{n-r} n$, for some sequences $L$, $R_1$, \dots, $R_{n-r}$ (where, of course, the $R_i$'s do not contain left-to-right maxima and $Ll$ is empty in case $\pi_1 =l$).
Then two distinct preimages of $\pi$ are $\pi_1 r L l M (r+1) R_1 (r+2) R_2 (r+3)\cdots n R_{n-r}$ and $r \pi_1 L l M (r+1) R_1 (r+2) R_2 (r+3)\cdots n R_{n-r}$.

The only remaining case is when $LTR(\pi)=\{n-k,\dots,n\}$ for some $k$, but some left-to-right maxima are adjacent. Let $i$ be the maximum integer such that $\pi_i=n-j \in LTR(\pi)$ and $\pi_{i+1}=n-j+1 \in LTR(\pi)$. Then $\pi$ can be written as $\pi= \pi_1 L (n-j) (n-j+1) R_{j-1} \cdots (n-1) R_1 n$ for some sequences $L$, $R_{j-1}$,\dots, $R_1$ (and the $R_i$'s as above and, again, $L(n-j)$ is empty in case $\pi_1 =n-j$). Then two distinct preimages of $\pi$ are $\pi_1 (n-j+1) L (n-j) (n-j+2) R_{j-1} \cdots n R_1$ and $(n-j+1) \pi_1 L (n-j) (n-j+2) R_{j-1} \cdots n R_1$.

We can now prove that $c_n^{(1)} = \sum_{k=2}^{\lceil\frac{n}{2}\rceil} (n-k)! \binom{n-k-1}{k-2}$. To obtain a permutation with exactly one preimage, we need to choose $k$ left-to-right maxima, which are forced to be the $k$ largest elements of the permutation, as they must be consecutive (warning: this is not the same $k$ as in the definition of $C_n^{(1)}$!). Since we require any two left-to-right maxima to be nonadjacent, then necessarily $k\leq \lceil \frac{n}{2} \rceil$. Moreover, any permutation with at least two elements and ending with its maximum has at least two left-to-right maxima, so $k\ge 2$. Once $k$ has been chosen, we need to choose the $k$ nonadjacent positions to place the left-to-right maxima (obviously in increasing order). Since the first and the last element of the permutations are necessarily left-to-right maxima (the last element in particular is $n$), the remaining $n-2$ elements consist of $k-2$ nonadjacent left-to-right maxima and $(n-2)-(k-2)=n-k$ non-left-to-right maxima. The possible ways to arrange these elements can be counted by placing the $n-k$ non-left-to-right maxima, then choosing $k-2$ inner spaces between them (among the $n-k-1$ available), which can be done in $\binom{n-k-1}{k-2}$ ways.  
Finally, the remaining elements of the permutation are smaller than all the left-to-right maxima and can be arranged in all possible ways, which are $(n-k)!$.

Summing up, we obtain $\sum_{k=2}^{\lceil\frac{n}{2}\rceil} (n-k)! \binom{n-k-1}{k-2}$ distinct permutations.
\end{proof}

The first terms of the sequence $c_n^{(1)}$ (starting from $n=1$) are $1,0,1,2,8,36,198,\ldots$, and are now recorded as sequence A374162 in the OEIS.

\begin{prop}
For $n\geq 4$, the set $C_n^{(2)}$ consists of all permutations of size $n$ ending with $n$ whose left-to-right maxima are consecutive and nonadjacent	\emph{except for the first one}, which \emph{is required} to be nonconsecutive with the second one, and \emph{can possibly} be adjacent to the second one. More formally,	
\begin{gather*}
C_n^{(2)}=\{\pi=\pi_1\cdots\pi_n\mid \pi_n=n \text{, there exists $k\ge 0$ such that } LTR(\pi)=\{\pi_1, n-k,\dots,n\} ,\\
\text{ with } \pi_1\neq n-k-1, \text{ and for every $\pi_i$, $\pi_j \in LTR(\pi)$, $\pi_i\neq\pi_j$, $i,j\neq 1$, we have } |j-i|>1 \}.
\end{gather*}

Moreover
\begin{equation}\label{two_preimages}
c_n^{(2)} = \sum_{k=3}^{n} \sum_{j=1}^{n-k} \frac{n-k-j+1}{j}(n-k)!\binom{n-j-k}{k-3}.
\end{equation}
\end{prop}
\begin{proof}
Let $\pi=\pi_1\cdots\pi_n$ be a permutation such that $\pi_n=n$, there exists $k\ge 0$ such that $LTR(\pi)=\{\pi_1, n-k,\dots,n\}$, with $\pi_1 \neq n-k-1$, and $\pi$ has no adjacent left-to-right maxima, except possibly for the first two. Thus we can write $\pi$ as $\pi=\pi_1 A_0 (n-k) A_k (n-k+1) A_{k-1}\cdots (n-1) A_{1} n$, and we know from \cref{preimages} that the only two preimages of $\pi$ are $\pi_1 (n-k) A_0 (n-k+1) A_k (n-k+2) A_{k-1}\cdots n A_{1}$ and $(n-k) \pi_1 A_0 (n-k+1) A_k (n-k+2) A_{k-1}\cdots n A_{1}$.

On the other hand, we now show that any permutation having a different form cannot have exactly two preimages. We can assume that $\pi$ is not as in the previous two propositions, otherwise it would have less than two preimages.
Similarly to what we did in the previous proposition, let $\pi=\pi_1\cdots \pi_n$ be a permutation with $\pi_n = n$ such that there exist $l,r\in LTR(\pi )$ that identify the rightmost gap in $LTR(\pi )$ \emph{excluding the gap identified by the first and the second left-to-right maxima}. 
So we can write $\pi$ as $\pi_1 L l M r R_1 (r+1) R_2 (r+2)\cdots R_{n-r} n$, for some sequences $L$, $R_1$, \dots, $R_{n-r}$. Notice that $L$ may be empty, but certainly $\pi_1\neq l$. Then $\pi$ has at least three different preimages, namely $\pi_1 r L l M (r+1) R_1 (r+2) R_2 (r+3)\cdots n R_{n-r}$, $r \pi_1 L l M (r+1) R_1 (r+2) R_2 (r+3)\cdots n R_{n-r}$ and $\alpha r M (r+1) R_1 (r+2) R_2 (r+3)\cdots n R_{n-r}$, where $\alpha$ is a preimage of the sequence $\pi_1 L l$ (this sequence has at least one preimage because it ends with its maximum).

The only remaining case occurs when $LTR(\pi)=\{\pi_1, n-k,\dots,n\}$ for some $k$ (with $\pi_1 \neq n-k-1$), and there is at least one pair of adjacent left-to-right maxima not involving $\pi_1$. Let $i>1$ be the maximum index such that $\pi_i=n-j$ and $\pi_{i+1}=n-j+1$ are both left-to-right maxima. Then $\pi$ can be written as $\pi= \pi_1 L (n-j) (n-j+1) R_{j-1} \cdots (n-1) R_1 n$ for some sequences $L$, $R_{j-1}$,\dots, $R_1$, with $L$ possibly empty. Then $\pi$ has at least three distinct preimages, namely $\pi_1 (n-j+1) L (n-j) (n-j+2) R_{j-1} \cdots n R_1$, $(n-j+1) \pi_1 L (n-j) (n-j+2) R_{j-1} \cdots n R_1$ and $\alpha (n-j+1) (n-j+2) R_{j-1} \cdots n R_1$, where $\alpha$ is a preimage of the sequence $\pi_1 L (n-j)$ (this sequence has at least one preimage because it ends with its maximum).

We are now left to prove the enumerative formula. We choose the number $k$ of left-to-right maxima of $\pi \in C_n^{(2)}$, and we observe that $k\geq 3$ (just as a consequence of the characterization of $C_n^{(2)}$ we have proved right above). 
We express $\pi$ as $\pi =L (n-k+2) R_1 \cdots (n-1) R_{k-2} n$, where $L$ is the longest prefix of $\pi$ containing no left-to-right maxima other than $\pi_1$. 
Denote with $j$ the length of $L$, and observe that $1\le j\le n-k$ (this is due to the fact that $L$ does not contain the largest $k-1$ left-to-right maxima and not even $n-k+1$). Since $\pi_1\le n-k$ for the non-consecutivity condition of $\pi_1$ and $n-k+2$, the elements of $L$ are all smaller than $n-k+1$. Therefore, we can choose the $j$ elements of $L$ in $\binom{n-k}{j}$ different ways. The maximum of the chosen elements is $\pi_1$, while the remaining $j-1$ elements can be permuted in all possible ways to obtain $L$. For the remaining part we reason in the same way of \cref{preim1} to obtain that we have $(n-j-k+1)! \binom{n-j-k}{k-3}$ different ways to choose a sequence with $n-j$ elements of which $k-1$ are non-adjacent left-to-right maxima. Summing everything together, we obtain that the number of permutations with exactly two preimages is given by (\ref{two_preimages}).
\end{proof}

\section{Preimages of classes}\label{classes}

The investigation of the preimages of a sorting algorithm is not limited to preimages of a single permutation. A common task in this framework is to study the preimages of sets of permutations, typically classes of permutations (i.e. sets of permutations defined in terms of the avoidance of certain patterns). Clear evidence for this approach exists in the literature, for instance in the cases of \texttt{Stacksort} \cite{D} (and references therein), \texttt{Queuesort} \cite{M} and \texttt{Bubblesort} \cite{AABCD}. When the preimages of permutation classes turn out to be classes themselves, the best way to describe them is to find an explicit basis. In case this is not possible, some kind of algorithmic procedure to determine the elements of the preimage is a notable alternative.

In the present section, we consider the preimages of principal classes under $\psb$, and we completely determine the permutations $\rho$ for which $\psb^{-1}(\Av (\rho ))$ is a class. Moreover, in all cases, we explicitly describe the basis of the resulting class. Our results will be useful in the next section, where we will compose \textsf{PSB} with other sorting algorithms.   

\bigskip

Our first result concerns the principal classes whose basis is a permutation that begins with its maximum.

\begin{prop}
\label{classPreimMax}
Given $n\geq 2$, let $\rho\in S_n$ be such that $\rho=n\alpha$, for some $\alpha \in S_{n-1}$. Then $\psb^{-1}(\Av(\rho))=\Av(B)$, where $B=\{ n(n+1)\alpha \} \cup \{ (n+2)n\tau \, |\, \tau \in (n+1)\shuffle \alpha , \tau \neq (n+1)\alpha \}$.
\end{prop}
\begin{proof}
Let $\rho=n\alpha$, with $\alpha \in S_{n-1}$. We start by proving that if a permutation $\sigma$ contains the pattern $\rho$, then for every permutation $\pi$ such that $\psb(\pi)=\sigma$, $\pi$ contains the pattern $n(n+1)\alpha$ or the pattern $(n+2)n\tau$, for some $\tau\in (n+1)\shuffle \alpha$, $\tau\neq (n+1)\alpha$.

Let $\pi$ be a permutation such that $\psb(\pi)=\sigma\ge\rho$. Then $\sigma$ contains an occurrence $m\beta$ of $\rho$, in which $m$ is larger than every element of the sequence $\beta$. By \cref{beta}, this implies that $\pi$ contains the subsequence $m\beta$ as well. We know that, during the sorting process, $m$ is sent to the output before the first element of $\beta$ becomes the current element of the input (notice that $\beta$ is not empty, because $n\geq 2$).  
There are two possible cases that lead to such a situation. 

If $m$ enters the pop stack, then there must exist an element $M$ larger than $m$ between $m$ and $\beta$ in $\pi$ (otherwise the pop stack would not be popped). Thus, $\pi$ contains the sequence $mM\beta$, which is an occurrence of the pattern $n(n+1)\alpha$.

On the other hand, suppose that $m$ does not enter the pop stack. Then, when $m$ is the current element of the input, there must be a left-to-right maximum $N$ of $\pi$, $N>m+1$, that appears before $m$ in $\pi$. Also, $\pi$ cannot contain the subsequence $N$, $N-1$,\dots, $m+2$, $m+1$, $m$, otherwise $m$ would be pushed into the pop stack. This implies that either there exists an element $M$, with $m<M<N$, which is not between $N$ and $m$ in $\pi$, or there exist two elements $M_1$, $M_2$, with $m<M_1<M_2<N$, such that $\pi$ contains the subsequence $N M_1 M_2 m$.

In the first case, we obtain that either $\pi$ contains the subsequence $MN\beta$, which is an occurrence of the pattern $n(n+1)\alpha$ in $\pi$, or $M$ is somewhere between two elements of $\beta$ or to the right of $\beta$ in $\pi$. In the latter case, the elements $N$, $m$, $\beta$ and $M$ form an occurrence of the pattern $(n+2)n\tau$, for some $\tau\in (n+1)\shuffle\alpha$, $\tau\neq (n+1)\alpha$.

In the second case, $M_1 M_2 \beta$ is an occurrence of $n(n+1)\alpha$ in $\pi$.

To conclude the proof, we show that, for every permutation $\pi$ containing one of the patterns in the statement of the proposition, $\psb(\pi)$ contains $\rho$.

Let $\pi$ be a permutation that contains an occurrence $M N \beta$ of the pattern $n(n+1)\alpha$. Then $\psb(\pi)$ contains the subsequence $M\beta$, since $M$ cannot be in the pop stack after $N$ has been processed by the algorithm (since $N>M$). Therefore, $\psb(\pi )$ contains an occurrence $M\beta$ of the pattern $n\alpha$.

On the other hand, suppose that $\pi$ contains an occurrence $NM\psi$ of a pattern $(n+2)n\tau$, for some $\psi\in M_1\shuffle\beta$, $\psi\neq M_1 \alpha$ (with $M<M_1 <N$). Note that $M$ is not part of a subsequence of consecutive elements of $\pi$ that begins with a left-to-right maxima, therefore, by \cref{LTRinStack}, $M$ does not enter the pop stack and $\psb(\pi)$ contains an occurrence $M\beta$ of $\rho$.
\end{proof}

Notice that, for $n=2$ and $\alpha=1$, the previous proposition gives $\psb^{-1}(\Av(21))=\Av(231,4213)$, which agrees with our characterization of sortable permutations provided in \cref{sortable}.

\bigskip

We next determine the preimage of a principal class whose basis is a permutation ending with its maximum and beginning with its second maximum.

\begin{prop}
Given $n\geq 3$, let $\rho\in S_n$, with $\rho=(n-1)\alpha n$, for some $\alpha \in S_{n-2}$. Then $\psb^{-1}(\Av(\rho))=\Av(B)$, where $B=\{ (n-1)n\alpha \} \cup \{ (n+1)(n-1)\tau \, |\, \tau \in n\shuffle \alpha , \tau\neq n\alpha \}$.
\end{prop}

\begin{proof}
Let $\rho=(n-1)\alpha n\in S_n$. We start by proving that if a permutation $\sigma$ contains the pattern $\rho$, then every permutation $\pi$ such that $\psb(\pi)=\sigma$ contains the pattern $(n-1)n\alpha$ or the pattern $(n+1)(n-1)\tau$, for some $\tau\in n\shuffle\alpha$, $\tau\neq n\alpha$.

Let $\pi$ be a permutation such that $\psb (\pi)=\sigma\ge\rho$. Then $\sigma$ contains an occurrence $m\beta M$ of $\rho$, with $m<M$ and $m>b$, for every element $b$ of the sequence $\beta$. By \cref{beta}, this implies that $\pi$ contains the subsequence $m\beta$ as well, however we do not have any information on the position of $M$ in $\pi$, so we examine all possible cases.

If $M$ is between $m$ and $\beta$, then $mM\beta$ forms an occurrence of $(n-1)n\alpha$ in $\pi$. 

If $M$ appears before $m$ in $\pi$, then $M$ is necessarily pushed into the pop stack, while $m$ is forced to bypass. Indeed, $M$ and $m$ appears in the output in reverse order, so necessarily $M$ cannot bypass; moreover, if $m$ were pushed into the pop stack while $M$ is still inside, then $\beta$ (which is not empty, since $n\geq 3$) would not appear in the output between $M$ and $m$. This means that $M$ and $m$ cannot be inside the pop stack together, and $M$ remains inside the pop stack until all the elements of $\beta$ are bypassed, so also $m$ is forced to bypass.  
As a consequence, by \cref{LTRinStack}, $m$ is not part of a sequence of consecutive elements of $\pi$ starting with a left-to-right maximum. Specifically, either there exists an element $M_1$, with $m<M_1<M$, which is not between $M$ and $m$ in $\pi$, or there exist two elements $M_1$, $M_2$, with $m<M_1<M_2<M$, such that $\pi$ contains the subsequence $M M_1 M_2 n$.
In the former case, either $\pi$ contains the subsequence $M_1 M\beta$, which is an occurrence of the pattern $(n-1)n\alpha$, or $M_1$ is somewhere between two elements of $\beta$ or to the right of $\beta$ in $\pi$. Thus the elements $M$, $m$, $\beta$ and $M_1$ form an occurrence of the pattern $(n+1)(n-1)\tau$, for some $\tau\in n\shuffle\alpha$, $\tau\neq n\alpha$.
In the latter case, $M_1 M_2 \beta$ is an occurrence of $(n-1)n\alpha$ in $\pi$.

Finally, we consider the case in which $M$ appears after the first element of $\beta$ in $\pi$. Then we consider $m$ and $\beta$ and argue in the same way as the proof of \cref{classPreimMax}, by distinguishing two cases, depending on whether $m$ enters the pop stack or not. In the former case we obtain an occurrence of $(n-1)n\alpha$ in $\pi$, whereas in the latter case we obtain either the pattern $(n-1)n\alpha$ or a sequence of elements that, together with $M$, forms an occurrence of a pattern $(n+1)(n-1)\tau$, for some $\tau\in n\shuffle\alpha$, $\tau\neq n\alpha$.

To conclude the proof, we show that, for every permutation $\pi$ containing one of the patterns in the statement of the proposition, $\psb(\pi)$ contains $\rho$.

Let $\pi$ be a permutation containing an occurrence $m M \beta$ of the pattern $(n-1)n\alpha$. Then $\psb(\pi)$ contains the subsequence $m\beta$, since $m$ cannot be in the pop stack after $M$ has been processed by the algorithm (because $M>m$).
Since \textsf{PSB} puts the maximum $N$ of $\pi$ at the end of the output permutation, we get that $\psb (\pi)$ contains an occurrence $m\beta N$ of the pattern $(n-1)\alpha n$ . Notice that $N$ might coincide with $M$, but it cannot be any of the elements of $\beta$, nor $m$.

On the other hand, suppose that $\pi$ contains an occurrence $M_1 m\psi$ of a pattern $(n+1)(n-1)\tau$, for some $\psi\in M\shuffle\alpha$, $\psi\neq M\alpha$. Note that $m$ is not part of a subsequence of consecutive elements of $\pi$ starting with a left-to-right maximum, since $\psi$ contains the element $M$, with $m<M<M_1$. Therefore, by \cref{LTRinStack}, $m$ is not pushed into the pop stack, hence $\psb(\pi)$ contains the subsequence $m\beta$. Since \textsf{PSB} puts the maximum $N$ of $\pi$ at the end of the output permutation, we have that $\psb(\pi )$ contains an occurrence $m\beta N$ of the pattern $(n-1)\alpha n$. Notice that $N$ might coincide with $M_1$, but it cannot be any of the elements of $\beta$, nor $m$ or $M$.
\end{proof}

We observe that for $n=3$ and $\alpha=1$, the previous proposition gives $\psb^{-1}(\Av(213))=\Av(231,4213)=\psb^{-1}(\Av(21))$. This is not surprising, since $\psb(\pi)$ ends with its maximum for all $\pi$. Therefore, if $\psb(\pi)$ contains an inversion, then it also contains the pattern $213$.

\bigskip

The patterns $\rho$ of the form $n \alpha$ and $(n-1)\alpha n$ are the only ones for which $\psb^{-1}(\Av(\rho))$ is a class. The next three propositions are devoted to show that this is indeed the case. In all cases, in order to prove that $\psb^{-1}(\Av(\rho))$ is not a permutation class, we exhibit two permutations $\sigma$, $\pi$ such that $\sigma\le\pi$ and $\rho\le\psb(\sigma)$, but $\rho\nleq\psb(\pi)$.

\begin{prop}
\label{noClassMiddle}
Let $\rho =\alpha n \beta\in S_n$, for some nonempty sequences $\alpha$, $\beta$. Then $\psb^{-1}(\Av(\rho))$ is not a permutation class.
\end{prop}
\begin{proof}
Let $\sigma = (n+1) \alpha (n+2) \beta n$ and $\pi=(n+1) (n+3) \alpha (n+2) \beta n$. Then $\sigma\le\pi$, because we obtain $\sigma$ by removing $n+3$ from $\pi$. Also, $\psb(\sigma) = \alpha (n+1) \beta n (n+2)$, which contains the occurrence $\alpha (n+1) \beta$ of $\alpha n \beta$.
On the other hand, $\psb(\pi)=(n+1)\alpha\beta n (n+2) (n+3)$, and we now show that such a permutation does not contain $\rho$. Indeed, since $\alpha$ and $\beta$ are not empty, $\rho$ neither begins nor ends at its maximum. Thus $n+3$ cannot be part of an occurrence of $\rho$ in $\psb(\pi)$. For the same reason, neither $n+2$ nor $n+1$ can. Therefore, an occurrence of $\rho$ in $\psb(\pi)$ should be an occurrence of $\rho$ in $\alpha\beta n$, which is impossible since $\rho$ and $\alpha\beta n$ have the same length and are different. Therefore $\rho\nleq\psb(\pi)$.
\end{proof}

\begin{prop}
Let $\rho =\alpha (n-1) \beta n\in S_n$, for some nonempty sequences $\alpha$, $\beta$. Then $\psb^{-1}(\Av(\rho))$ is not a permutation class.
\end{prop}
\begin{proof}
Let $\sigma = n \alpha (n+1) \beta (n-1)$ and $\pi=n (n+2) \alpha (n+1) \beta (n-1)$. Then $\sigma\le\pi$, because we obtain $\sigma$ by removing $n+2$ from $\pi$. Also, $\psb(\sigma) = \alpha n \beta (n-1) (n+1)$, which contains the occurrence $\alpha n \beta (n+1)$ of $\rho$.
On the other hand, $\psb(\pi)=n\alpha\beta (n-1) (n+1) (n+2)$, and we now show that such a permutation does not contain $\rho$. Indeed, if we remove both $n+1$ and $n+2$ from $\psb(\pi)$, the resulting permutation (of length $n$) is different from $\rho$, therefore, to have an occurrence of $\rho$ in $\psb(\pi)$, at least one of $n+1$ and $n+2$ must be involved. Actually, we cannot choose both, since the maximum and the second maximum are not adjacent in $\rho$. Therefore $\psb(\pi)$ contains an occurrence of $\rho$ if and only if $\pi'=n\alpha\beta (n-1)$ contains an occurrence of $\rho'=\alpha (n-1) \beta$. Since $\alpha$ and $\beta$ are not empty, $\rho'$ neither begins nor ends with its maximum $n-1$. Hence $n$ and $n-1$ cannot be part of an occurrence of $\rho'$ in $\pi'$, but $\rho'$ has length $n-1$ and $\pi'$ has length $n$. Therefore, there are no occurrences of $\rho'$ in $\pi'$, that is, $\rho\nleq\psb(\pi)$.
\end{proof}

\begin{prop}
Let $\rho =\alpha (n-1) n\in S_n$, with $\alpha\in S_{n-2}$ and $n\ge 4$. Then $\psb^{-1}(\Av(\rho))$ is not a permutation class.
\end{prop}
\begin{proof}
Let $\sigma = (n-1) (n+1) n \alpha$ and $\pi = (n-1) (n+1) (n+2) n \alpha$. Then $\sigma\le\pi$ because we obtain $\sigma$ by removing $n+2$ from $\pi$.
Also, $\psb(\sigma) = (n-1) \alpha n (n+1)$, which contains the occurrence $\alpha n (n+1)$ of $\rho$.
On the other hand, $\psb(\pi)=(n-1) (n+1) n \alpha (n+2)$, and we now show that such a permutation does not contain $\rho$. Indeed, since $\alpha$ has at least two elements and $(n-1)(n+1)n(n+2)$ cannot be an occurrence of $\rho$, then a possible occurrence of $\rho$ in $\psb(\pi)$ should contain at least one element of $\alpha$.
Moreover, since $\rho$ ends with its maximum, $n+2$ should belong to such an occurrence. Finally, since $\rho$ has $n$ elements, at least one element amongst $n-1$, $n+1$ and $n$ should be selected to form our occurrence. However, this is impossible, because the three latter elements are all greater than each element of $\alpha$, and $\rho$ ends with its two largest elements. Therefore, $\rho\nleq\psb(\pi)$.
\end{proof}

We remark that, if $\rho=123$, we cannot use the same argument as in the above proof to conclude that $\psb^{-1}(\Av(\rho))$ is not a permutation class. This is however true, by choosing the two permutations $\sigma=321$ and $\pi=3421$.

\section{Compositions with other sorting algorithms}\label{composition}

If we execute a sorting algorithm $\mathbf{X}$ on a permutation and then execute another sorting algorithm $\mathbf{Y}$ on the output permutation, we get a new sorting algorithm, whose associated map is the \emph{composition} of the maps associated with $\mathbf{X}$ and $\mathbf{Y}$.   

In this section, we investigate the permutations that can be sorted using a composition of \textsf{PSB} with another sorting algorithm. Specifically, we will consider the algorithms \texttt{Queuesort}, \texttt{Stacksort} and \texttt{Bubblesort}, whose associated maps will be denoted $\que$, $\stack$ and $\bub$, respectively. In several cases, we obtain potential matches with sequences in \cite{Sl}, which we propose as conjectures for possible further work. 

\bigskip

We begin by considering compositions in which \textsf{PSB} is performed first.

\begin{prop}
\label{stackPopstack}
Let $\pi\in S_n$. Then $\stack\circ\psb(\pi)=id_n$ if and only if $\pi\in\Av(2341, 25314, 52314,$ $45231,$ $42531,3\bar{5}241)$. 
\end{prop}

\begin{proof}
The algorithm \texttt{Stacksort} sorts a permutation if and only if it belongs to the permutation class $\Av(231)$. Therefore, $\pi$ is not sortable by the composition $\stack \circ \psb$ if and only if $\psb(\pi)$ contains the pattern $231$. 

We start by supposing that $\pi$ is a permutation which is not sortable by $\stack\circ\psb$. Then there exist three elements $a$, $b$, $c$, with $a<b<c$, such that $\psb(\pi)$ contains such elements in the order $b,c,a$. By \cref{noInv}, we have that $\pi$ contains the elements $a$, $b$, $c$, either in the order $b$, $c$, $a$ or in the order $c$, $b$, $a$.

Suppose that $\pi$ contains the elements $b$, $c$, $a$ in this order. We know that, when we apply the algorithm \textsf{PSB}, these elements remain in the same order. Suppose that $c$ is pushed into the pop stack during the sorting process. Then $c$ must reach the output before $a$ is the current element of the input, otherwise $a$ would bypass, thus reaching the output before $c$. Therefore there must be an element $d$ between $c$ and $a$ in $\pi$ that forces the pop stack to be emptied. Since this happens only when the current element of the input is larger than the content of the pop stack, we have $d>c$, hence $\pi$ contains the pattern $2341$ (given by the elements $b$, $c$, $d$, $a$). On the other hand, suppose that $c$ is not pushed into the pop stack during the sorting process. Then it is not a left-to-right maximum, and there exists at least one element larger than $c$ and to the left of $c$ in $\pi$. Let $e$ be the rightmost of such elements. Now consider the instant in which $c$ is the current element of the input. By construction, $e$ is in the pop stack. Since $c$ does not enter the pop stack by hypothesis, then $\pi$ does not contain the decreasing sequence $e$, $e-1$, \dots, $c+1$, $c$. Therefore, either there exists an element $d$, with $c<d<e$, which is not between $e$ and $c$ in $\pi$, or there exist two elements $d_1$ $d_2$, with $c<d_1<d_2<e$, that appear in $\pi$ in the order $e$, $d_1$, $d_2$, $c$. The first case gives one of the patterns 2341, 25314, 42531, 45231 or 52314, depending on the positions of $d$ and $e$ with respect to $b$, $c$, and $a$. The second case gives one of the patterns 2341, 45231 or 42531, depending on the positions of $d_1$ and $d_2$ with respect to $b$, $c$, and $a$.

On the other hand, suppose that $\pi$ contains the elements $c$, $b$, $a$ in this order, and during the sorting process $b$ reaches the output before $c$. For this to happen, $c$ must be pushed into the pop stack and stay there until $b$ becomes the current element. Moreover, the pop stack must be emptied before $a$ becomes the current element. This can occur only if there is an element $d$ greater than $c$ between $b$ and $a$ in $\pi$. Note that the elements $c$, $b$, $d$, $a$ form an occurrence of the pattern 3241 in $\pi$. Since $c$ must remain in the pop stack until $b$ is the current element of the input, there cannot be any element larger than $c$ between $c$ and $b$ in $\pi$. This implies that $\pi$ must avoid the patterns $3\bar{5}241$ and $3\bar{4}251$. However, we can ignore the pattern $3\bar{4}251$, since an occurrence of the pattern 34251 would contain an occurrence of the pattern 2341, which is prohibited.

In order to prove the converse statement, we need to show that, if a permutation $\pi$ contains one of the patterns specified in the statement, then $\stack\circ\psb(\pi)$ is not the identity permutation. Equivalently, this amounts to prove that $\psb(\pi)$ contains 231. This can be done by means of a (somehow lengthy but) rather easy case-by-case analysis.  
\end{proof}

Notice that the set of permutations such that $\stack\circ\psb(\pi)=id_n$ is not a permutation class (this is due to the presence of a barred pattern among those to be avoided). This is consistent with \cref{noClassMiddle}, which asserts that $\psb^{-1}(\Av(231))$ is not a class. For this reason, the previous proposition needed an \emph{ad-hoc} proof.
This is not the case for \texttt{Queuesort}, which can be more easily dealt with using the tools developed in the previous section.

\begin{prop}
\label{quePopstack}
Let $\pi\in S_n$. Then $\que\circ\psb(\pi) = id_n$ if and only if $\pi\in\Av(3421, 53241, 53214)$.
\end{prop}

\begin{proof}
The algorithm \texttt{Queuesort} sorts a permutation if and only if it belongs to the permutation class $\Av(321)$.
Therefore, the set of permutations that can be sorted by the above composition is $\psb^{-1}(\Av(321))$.  
We can thus invoke \cref{classPreimMax} to obtain $\psb^{-1}(\Av(321))=\Av(3421, 53241, 53214)$, which concludes the proof.
\end{proof}

The first terms of the sequence enumerating the above class are $1,2,6,23,101,480,2400,$ $12434,66142,359112,1981904,11085198,\ldots$, and seem to match sequence A218225 in \cite{Sl}, which is also conjectured to enumerate different classes of pattern avoiding permutations.

Finally, we consider the composition of $\psb$ with $\bub$. Since the set of permutations sortable using \texttt{Bubblesort} is characterized by the avoidance of the two patterns $231$ and $321$, we can use the previous propositions to obtain the set of sortable permutations.

\begin{prop}
\label{bubPopstack}
Let $\pi\in S_n$. Then $\bub\circ\psb(\pi)=id_n$ if and only if $\pi\in\Av(2341, 3421, 3241,$ $25314, 52314, 53214)$.
\end{prop}

\begin{proof}
The algorithm \texttt{Bubblesort} sorts a permutation if and only if it belongs to the permutation class $\Av(231, 321)$. Therefore, $\pi$ is not sortable by the composition of $\bub \circ \psb$ if and only if $\psb(\pi)$ contains the pattern $231$ or the pattern $321$. 
Since such patterns are those considered in \cref{stackPopstack} and \cref{quePopstack}, we can intersect the corresponding preimages,
that is,
\begin{align*}
\psb^{-1}(\Av(231,321)) =& \psb^{-1}(\Av(231))\cap\psb^{-1}(\Av(321))\\
=& \Av(2341, 25314, 52314, 45231, 42531, 3\bar{5}241) \cap \Av(3421, 53241, 53214)\\
=& \Av(2341, 25314, 52314, 45231, 42531, 3\bar{5}241, 3421, 53241, 53214).
\end{align*}

Notice that a permutation avoids the patterns 3421 and $3\bar{5}241$ if and only if it avoids the patterns 3421 and 3241. Also, the avoidance of the patterns 45231, 42531 and 53241 is implied by the avoidance of the patterns 3421 and 3241. Therefore

\begin{align*}
\psb^{-1}(\Av(231,321)) =& \Av(2341, 25314, 52314, 45231, 42531, 3\bar{5}241, 3421, 53241, 53214)\\
=& \Av(2341, 3421, 3241, 25314, 52314, 53214).
\end{align*}

\end{proof}

The inverse of the above class (i.e. the class whose basis is the set of inverses of the above permutations) has a regular insertion encoding \cite{V}, thus it is possible to automatically deduce its rational generating function \cite{ABCNPU, BEMNPU}, which is
\[
\frac{(3x-1)(x^2 +2x-1)(x-1)^2}{5x^5 +6x^4 -24x^3 +22x^2 -8x+1}.
\]

The associated coefficients start $1,2,6,21,76,273,970,3422,12027,42194,147901,518206,\ldots$ and do not appear in \cite{Sl}.

\bigskip

We now consider compositions of $\psb$ with the same algorithms as above, but this time we choose to perform \textsf{PSB} last. In our first result, to find the basis of the preimage we exploit some properties of \texttt{Queuesort}, for which we refer the reader to \cite{M}.

\begin{prop}
\label{popQueue}
Let $\pi\in S_n$. Then $\psb \circ \que (\pi ) = id_n$ if and only if $\pi \in \Av (4231, 2431, 54213)$.
\end{prop}
\begin{proof}
The algorithm \textsf{PSB} sorts a permutation if and only if it belongs to the permutation class $\Av(231,4213)$. Therefore, $\pi$ is not sortable by the composition of $\psb \circ \que$ if and only if $\que(\pi)$ contains the pattern $231$ or the pattern $4213$.

Suppose that $\que(\pi)$ contains the pattern $231$, and let $bca$ be one of its occurrences. Since \texttt{Queuesort} does not create new inversions, $\pi$ contains either the sequence $bca$ or the sequence $cba$. 

Assume that $\pi$ contains the sequence $bca$. If $c$ is not a left-to-right maximum of $\pi$, then there exists an element $d$ greater than $c$ to the left of $c$ in $\pi$. Therefore, $\pi$ contains the pattern $4231$ or the pattern $2431$. If $c$ is a left-to-right maximum, then it enters the queue during the sorting process. Since $c$ must be popped before $a$ is the current element, there must be an element $d$ before $a$ in $\pi$, such that $d$ is larger than $c$ but smaller than the last element $e$ of the queue. The elements $b$, $e$, $d$, $a$ form the pattern $2431$ in $\pi$.

On the other hand, assume that $\pi$ contains the sequence $cba$. Then $c$ must be a left-to-right maximum, otherwise $c$ and $b$ would also form an inversion in the output. Therefore, $c$ enters the queue and must exit before $a$ is the current element of the input. Therefore, there must be an element $d$ before $a$ in $\pi$, such that $d$ is greater than $c$ but smaller than the last element $e$ of the queue. The elements $c$, $e$, $d$, $a$ form the pattern $2431$ in $\pi$. This solves the case in which $\que(\pi)$ contains the pattern $231$.

Now suppose that $\que(\pi)$ contains the pattern $4213$, and let $dbac$ be one of its occurrences. Since \texttt{Queuesort} does not create new inversions, $\pi$ contains the sequence $dcba$, the sequence $dbca$ or the sequence $dbac$ (and of course only one case can occur). However, notice that the first two cases cannot occur, as in each of them $c$ is not a left-to-right maximum of $\pi$, so the inversion between $c$ and $a$ would remain in the output. Hence $\pi$ contains the sequence $dbac$. If $d$ is not a left-to-right maximum of $\pi$, then there exists an element $e$ greater than $d$ that forms the pattern $54213$ together with $d$, $b$, $a$ and $c$. Otherwise, $d$ is a left-to-right maximum of $\pi$ and enters the queue during the sorting of $\pi$ with \texttt{Queuesort}. Since $d$ goes to the output before $b$ is considered, there exists an element $e$ between $d$ and $b$ in $\pi$ such that $e>d$ and $e$ is smaller than the last element $f$ of the queue. The elements $d$, $f$, $e$ and $b$ form the pattern $2431$ in $\pi$. This concludes the proof.
\end{proof}

The first terms of the associated enumerating sequence are $1, 2, 6, 22, 89, 380, 1679, 7602,$ $35072, 164266, 779022, 3733444,\ldots$, and match the first terms of sequence A049123 in \cite{Sl}, for which no combinatorial interpretation is provided.

\begin{prop}
\label{popBubble}
Let $\pi\in S_n$. Then $\psb\circ\bub(\pi) = id_n$ if and only if $\pi\in\Av(2341,2431,3241,$ $4231,45213,54213)$.
\end{prop}
\begin{proof}
The algorithm \textsf{PSB} sorts a permutation if and only if it belongs to the permutation class $\Av(231, 4213)$. We can use the results of \cite{AABCD} to find the preimages $\bub^{-1}(\Av(231))$ and $\bub^{-1}(\Av(4213))$, and then intersect the two sets. They are both permutation classes, specifically $\bub^{-1}(\Av(231))=\Av(2431, 4231, 2341, 3241)$ and $\bub^{-1}(\Av(4213))=\Av(45213, 54213)$, which gives the thesis.
\end{proof}

Also in this case, as it happened for Proposition \ref{bubPopstack}, the (inverse of the) above class has a regular insertion encoding, so we can compute the associated generating function, which is
\[
\frac{1-3x}{1-4x+2x^2}.
\]

The sequence of the coefficients of the above generating function is (essentially) A006012 in \cite{Sl}. This is a very interesting sequence that satisfies a simple linear recurrence relation, with plenty of combinatorial interpretations. The present one can thus be added to the list. 

\section{Pop stacks in parallel with a bypass}\label{parallel}

As with other sorting devices, we can connect multiple pop stacks with a bypass option in different ways to create more robust sorting machines.  In this section, we consider one of the most natural such options by putting pop stacks in parallel where for each entry of the input, there is still an option to bypass the pop stacks and go directly to the output.  Pop stacks in parallel were originally studied by Atkinson and Sack~\cite{AS} who determined a recurrence relation for the enumeration of the sortable permutations in the case of two pop stacks in parallel; they also proved that the set of forbidden patterns that characterize the sortable permutations for any number $k$ of pop stacks in parallel is finite. Atkinson and Sack~\cite{AS} also conjectured that the generating function for the enumeration of sortable permutations for any value $k$ is rational, which was proved by the last author and Vatter~\cite{SV}.

As before, we begin with the necessary condition that sortable permutations must avoid a certain class of permutations.

\begin{prop}
    Let $\pi$ be a permutation that contains one of the following permutations: $$2341,25314,42513,42531,45213,45231,52314,642135,642153.$$  Then $\pi$ cannot be sorted by two pop stacks in parallel with a bypass operation.
\end{prop} 

\begin{proof}
    With each of the patterns $2341,25314,42513,42531,45213,45231,52314,642135,642153$, consider what must be done to sort a permutation that contains them.  Recall that, when sorting a permutation, the entries of a pop stack must be in increasing order from top to bottom and also be such that each entry $\pi_i$ is exactly one less than the entry immediately below $\pi_i$ in the pop stack.  
    
    Consider an attempt to sort any of the patterns by two pop stacks in parallel with a bypass.  Notice that, for any of the patterns $2341,25314,42513,42531,45213,45231,52314,$ $642135,642153$, at least the first three entries must enter a pop stack before any can be popped, as there is a later entry in the pattern that is smaller than all three. However, we see that the second entry, say $\pi_{\alpha_2}$, of each pattern must enter a different pop stack from that entered by the first entry, say $\pi_{\alpha_1}$, as either $\pi_{\alpha_2} > \pi_{\alpha_1}$ ($2341,25314,45213,45231$) or there is both a later entry, call it $\pi_{\alpha_k}$, such that $\pi_{\alpha_2} < \pi_{\alpha_k} < \pi_{\alpha_1}$ and so $\pi_{\alpha_2}$ cannot be one less than the current top element of the pop stack $\pi_{\alpha_1}$ entered ($42513,42531,52314,642135,642153$).  Finally, the third entry, say $\pi_{\alpha_3}$ of each pattern, can now not enter either of the pop stacks for the same reasons that $\pi_{\alpha_2}$ could not enter the same pop stack as $\pi_{\alpha_1}$.

    Hence, any algorithm will not be able to sort a permutation containing any of the patterns $2341,25314,42513,42531,45213,45231,52314,642135,642153$.
\end{proof}

Below we give Algorithm~\ref{PSBP}, which we will denote as \textsf{PSBP}. Like \textsf{PSB}, this algorithm chooses to push an entry into a pop stack instead of bypassing the stacks if both moves are an option to sort a permutation.  

\begin{algorithm}\label{PSBP}
	$S_1, S_2, O:=\emptyset$\;
	$i:=1$\;
	\While{$i\leq n$}
	{
        \If{$\pi_i = \textnormal{\texttt{TOP}}(S_1)-1$}
		{
				$\textnormal{\texttt{PUSH}}_1$\;	
		}
		\ElseIf{$\pi_i = \textnormal{\texttt{TOP}}(S_2)-1$}
		{
				$\textnormal{\texttt{PUSH}}_2$\;	
		}
        \ElseIf{$S_1 =\emptyset$}
		{
			$\textnormal{\texttt{PUSH}}_1$\;
        }
        \ElseIf{$S_2 =\emptyset$}
		{
			$\textnormal{\texttt{PUSH}}_2$\;
        }
		\ElseIf{$\pi_i < \max(\textnormal{\texttt{TOP}}(S_1)-1,\textnormal{\texttt{TOP}}(S_2)-1)$}
		{
			\texttt{BYPASS}\;
		}
		\Else
        {
            \If{$\textnormal{\texttt{TOP}}(S_1) < \textnormal{\texttt{TOP}}(S_2)$}
            {
                $\textnormal{\texttt{POP}}_1$\;
                $\textnormal{\texttt{PUSH}}_1$\;
            }
            \Else
            {
                $\textnormal{\texttt{POP}}_2$\;
                $\textnormal{\texttt{PUSH}}_2$\;
            }
		}
    }
    \If{$S_1 =\emptyset$}
	{
		$\textnormal{\texttt{POP}}_2$\;	
	}
    \ElseIf{$S_2 =\emptyset$}
    {
        $\textnormal{\texttt{POP}}_1$\;
    }
    \ElseIf{$\textnormal{\texttt{TOP}}(S_1) < \textnormal{\texttt{TOP}}(S_2)$}
    {
        $\textnormal{\texttt{POP}}_1$\;
        $\textnormal{\texttt{POP}}_2$\;
    }
    \Else
    {
        $\textnormal{\texttt{POP}}_2$\;
        $\textnormal{\texttt{POP}}_1$\;
    }
	\caption{\textsf{PSBP} ($S_j$ is pop stack $j$; $O$ is the output; \texttt{TOP}($S_j$) is the current top element of the pop stack $S_j$; 
 operations $\textnormal{\texttt{PUSH}}_j$, $\textnormal{\texttt{POP}}_j$, \texttt{BYPASS} are ``insert $\pi_i$ into $S_j$", ``pour the whole content of $S_j$ into the output", ``bypass the pop stacks", respectively;  
    $\pi=\pi_1 \cdots \pi_n$ is the input permutation).}
\end{algorithm}

We can show \textsf{PSBP} is optimal in the sense that it sorts all permutations sortable by the machine consisting of two pop stacks in parallel with a bypass option.  

\begin{prop}~\label{P:PSRB}
	The set of permutations sortable by \textsf{PSBP} 
    is the class
	$$\Av (2341,25314,42513,42531,45213,45231,52314,642135,642153).$$
\end{prop} 

\begin{figure}[t]
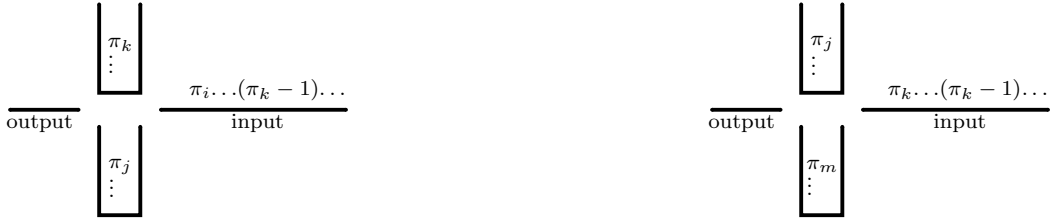

\begin{center}

\begin{tabular}{l r}
\begin{twostack}
\fillstacktwo{}{{.}{.}{.}{\pi_k}}{{.}{.}{.}{\pi_j}}{{\pi_i} {\ldots} {(\pi_{k}-1)}  {\ldots} }
\end{twostack}
\qquad \qquad \qquad
&
\qquad \qquad \qquad
\begin{twostack}
\fillstacktwo{}{{.}{.}{.}{\pi_j}}{{.}{.}{.}{\pi_m}}{{\pi_k} {\ldots} {(\pi_{k}-1)}  {\ldots} }
\end{twostack}
\\
\end{tabular}

\caption{The stage immediately prior to which \textsf{PSBP} first fails to sort a permutation $\pi$. The left picture shows when $\pi_k$ will enter the output via a pop stack and the right picture shows when $\pi_k$ will enter the output via the bypass operation. }
\label{fig_fail}
\end{center}
\end{figure}

\begin{proof}
    Consider a permutation $\pi$ that cannot be sorted by \textsf{PSBP}.  Look specifically at the critical stage at which \textsf{PSBP} fails in the sense that an entry $\pi_k$ is output before $\pi_k-1$ has been output (we may assume that this is the first failure and thus $\pi_k$ is the smallest entry to enter the output prematurely). Consider whether $\pi_k$ enters the output from a pop stack or $\pi_k$ passes to the output directly from the input via the bypass operation.  
    
    First suppose $\pi_k$ is popped from one of the pop stacks before $\pi_k - 1$ is output. Notice that \textsf{PSBP} operates in such a manner that the entries of each pop stack are always increasing and in consecutive order from top to bottom.  Thus $\pi_k$ must be the top entry of one of the pop stacks. Say $\pi_j > \pi_k$ is the top entry of the other pop stack.   Suppose the next entry from the input is $\pi_i$.  The stage immediately before $\pi_k$ is popped to the output is shown on in the left picture of Figure~\ref{fig_fail}.  Based on the algorithm, $\pi_i > \pi_j$ or $\pi_k < \pi_i < \pi_j -1$.   Also, $\pi_k -1$ must appear after $\pi_i$ in $\pi$.  
    
    In the instance $\pi_i > \pi_j$, consider whether $\pi_k$ or $\pi_j$ appeared earlier in $\pi$.  If $k<j<i$, then $\pi$ contains a pattern $\pi_k \pi_j \pi_i (\pi_k-1)$.  That is, $\pi$ contains $2341$.  Otherwise if $j<k<i$, then the entry $\pi_{j}-1$ must appear before $\pi_j$ in $\pi$ or after $\pi_k$ (and thus after $\pi_i$ as well) for $\pi_j-1$ to not be directly above $\pi_j$ in the pop stack.  Therefore, $\pi$ contains a pattern $(\pi_j -1) \pi_j \pi_i (\pi_k - 1)$ in the former case or one of the patterns $\pi_j \pi_k \pi_i (\pi_j-1) (\pi_k-1)$ or $\pi_j \pi_k \pi_i (\pi_k-1) (\pi_j-1)$ in the latter case.  That is, $\pi$ contains $2341, 42531,$ or $42513$.
    
    In the other instance where $\pi_k$ is popped to the output, we have $\pi_k < \pi_i < \pi_j -1$.  It must be the case that $\pi_j - 1$ appears after $\pi_i$ as $\pi_k < \pi_i < \pi_j -1$, so $\pi_j -1$ cannot be in the same pop stack as $\pi_k$, as the entries there are in consecutive order and $\pi_j - 1$ is too small to be in the pop stack with $\pi_j$ on top. If $k<j<i$, then $\pi$ contains a pattern $\pi_k \pi_j \pi_i (\pi_j-1) (\pi_k-1)$ which contains $2341$ or a pattern $\pi_k \pi_j \pi_i (\pi_k-1) (\pi_j-1)$ (which is $25314$).  If instead $j<k<i$, then $\pi$ contains $\pi_j \pi_k \pi_i (\pi_j-1) (\pi_k-1)$ which contains $2341$ or a pattern $\pi_j \pi_k \pi_i (\pi_k-1) (\pi_j-1)$ (which is $52314$). 
    
    The last possibility is that $\pi_k$ is output before $\pi_{k}-1$ through a bypass operation.  Due to the requirements of $\textsf{PSBP}$, this means $\pi_k -1$ appears after $\pi_k$ in $\pi$ and also both stacks have top entries $\pi_j, \pi_m >\pi_k$.  The stage immediately before $\pi_k$ is bypassed to the output is shown on in the right picture of Figure~\ref{fig_fail}. Without loss of generality, suppose $\pi_j<\pi_m$.  For $\pi_k$ not to have been pushed on top of $\pi_j$, there must be another entry $\pi_j - 1$ that appears after $\pi_k$ in $\pi$.  In the case where $j<m$, this means $\pi$ contains the pattern $\pi_j \pi_m \pi_k (\pi_j -1) (\pi_k - 1)$ or $\pi_j \pi_m \pi_k (\pi_k -1) (\pi_j - 1)$, that is $\pi$ contains $45231$ or $45213$.  Otherwise, if $m<j$, for $\pi_j$ to have not been pushed on top of $\pi_m$, we must have had the entry $\pi_m -1$ appear before $\pi_m$ or after $\pi_k$.  That is, $\pi$ contains one of the following patterns:
    
    $(\pi_m -1) \pi_m \pi_j \pi_k (\pi_j -1) (\pi_k - 1)$, i.e contains $45231$ (without $\pi_j$),

    $(\pi_m -1) \pi_m \pi_j \pi_k (\pi_k -1) (\pi_j - 1)$, i.e contains $45213$ (without $\pi_j$),
    
    $ \pi_m \pi_j \pi_k (\pi_m -1) (\pi_j -1) (\pi_k - 1)$, i.e contains $42531$ (without $\pi_m$),
    
    $ \pi_m \pi_j \pi_k (\pi_m -1) (\pi_k -1) (\pi_j - 1)$, i.e contains $42513$ (without $\pi_m$),
    
    $ \pi_m \pi_j \pi_k  (\pi_j -1) (\pi_m -1)(\pi_k - 1)$, i.e contains $2341$ (without $\pi_m,\pi_j$),
    
    $ \pi_m \pi_j \pi_k  (\pi_k -1) (\pi_m -1)(\pi_j - 1)$, i.e contains $642153$,
    
    $ \pi_m \pi_j \pi_k  (\pi_j -1) (\pi_k -1)(\pi_m - 1)$, i.e contains $52314$ (without $\pi_j$),
    
    $ \pi_m \pi_j \pi_k  (\pi_k -1) (\pi_j -1)(\pi_m - 1)$, i.e contains $642135$.
\end{proof}

Once again, the inverse of the above class has a regular insertion encoding \cite{V}, thus it is possible to automatically deduce its rational generating function \cite{ABCNPU, BEMNPU}, which is
\begin{equation}~\label{E:Parallel_Gen}
\frac{(1-x)(1-2x)(1-4x)}{1-8x+20x^2 -18x^3 +3x^4}.
\end{equation}

The sequence of coefficients has now been recorded as A374165 in \cite{Sl}.

In fact, we can show that any machine made up of $k$ pop stacks in parallel with a bypass option will sort a class of permutations characterized by the avoidance of a finite set of patterns.  This is an extension of work of Atkinson and Sack~\cite[Theorem 2]{AS} showing that a machine made up of $k$ pop stacks in parallel has sortable permutations being exactly a permutation class with a finite basis.  We can even utilize their constructive technique which is not dissimilar to specific case shown in Proposition~\ref{P:PSRB}.  
As there is a small error in the original published construction, we include the full proof of Theorem~\ref{T:finite_basis} here.  We also note that the basis given following Atkinson and Sack~\cite[Theorem 2]{AS} for the $k=2$ case is actually the complement of the correct basis. Indeed, the set of permutations sortable by two pop stacks in parallel (without a bypass) is $\Av{(2341,3412,25314,42531, 52314,53124,53142)}$.

Before stating Theorem \ref{T:finite_basis}, we will describe an algorithm that will sort a sortable permutation $\pi$  by $k$ pop stacks in parallel with a bypass where we do not concern ourselves with the image of any non-sortable permutation.  At each stage, the process (in priority order) is as follows:
\begin{enumerate}
\item If the next entry of the input $\pi_j$ is the next needed entry for the output, $\pi_j$ goes directly to the output via the bypass operation.
\item If the top entry of any pop stack $S_i$ is the next needed entry for the output, pop $S_i$.
\item If the next entry of the input $\pi_j$ is exactly one less than the top entry $t_i$ of pop stack $S_i$, push $\pi_j$ on top of $t_i$.
\item If the next entry of the input $\pi_j$ is not exactly one less than the top entry of any of the pop stacks, push $\pi_j$ into an empty pop stack.
\item If none of these moves are possible, $\pi$ is not sortable.
\end{enumerate}

With this set of legal moves in mind, we can now prove that the basis for the class of permutations sortable by $k$ pop stacks in parallel with a bypass is finite.

\begin{theorem}~\label{T:finite_basis}
    There is a finite set of permutations $B_k$ such that a permutation $\pi$ is sortable by $k$ pop stacks in parallel with a bypass if and only if $\pi \in \Av{(B_k)}$.
\end{theorem}

\begin{proof}
   	Consider a permutation $\pi$ that cannot be sorted by $k$ pop stacks in parallel with a bypass option by any algorithm.  Specifically, there must be a point of failure in any process where no legal move can be made.  At this stage,
    \begin{enumerate}
        \item every pop stack $S_i$ contains a nonempty consecutive sequence of entries increasing from top to bottom with top entry $t_i$ and bottom entry $b_i$,
        \item the last entry of the output is $c$ (where $c = 0$ if the output is currently empty),
        \item the current input entry is $\pi_j$
        \item $t_i > c +1$ for all $i$,
        \item For each $i$, we have either $\pi_j > t_i$ or $\pi_j < t_i - 1$, and
        \item $\pi_j > c+1$.
    \end{enumerate}  

    Notice that each entry $b_i$, if $b_i \neq n$, must have entered the stacks prior to $b_{i}+1$, as otherwise $b_i$ would have been placed on top of $b_{i}+1$ instead of being put at the bottom of an empty stack.  Further, the entry $c+1$ is not in the stacks, so must appear after $\pi_j$ in $\pi$.  Finally, $\pi_j$ cannot enter any stack, so if $\pi_j \neq n$, then $\pi_j+1$ must appear after $\pi_j$.

    However, conversely, suppose we have any set of (not necessarily distinct) entries \\
    $\{b_1,b_2, \ldots, b_k, b_1+1, b_2+1, \ldots, b_k+1, \pi_j, \pi_j + 1, c+1 \}$ (where we eliminate the largest entry) such that 
    \begin{enumerate}
        \item $b_i$ appears before $b_i + 1$,
        \item $\pi_j$ appears after $b_i$ for all $i$
        \item $\pi_j$ appears before $\pi_j + 1$, and
        \item $c+1$ appears after $\pi_j$.
    \end{enumerate}
    Then each $b_i$ needs to be pushed into a different pop stack, none of the pop stacks can be popped, and $\pi_j$ has no legal move (push or bypass).  Thus we have a finite set of permutations that can minimally fulfill the constraints forcing a failure.  Therefore the basis of permutations that can be sorted by a machine consisting of $k$ pop stacks in parallel with a bypass is finite.
\end{proof}

It is a straightforward extension of the proof of \cite[Theorem 1]{SV} to show that the generating function for the enumeration of permutations sortable by $k$ pop stacks in parallel with a bypass is rational for any $k$, so we include the following theorem without further justification.

\begin{theorem}
    For any positive integer $k$, the set of permutations sortable by $k$ pop stacks in parallel with a bypass has a rational generating function.
\end{theorem}

\section{Further work}\label{final}

\subsection{An alternate algorithm and sorting words}

While \textsf{PSB} is an ideal algorithm to handle permutations when sorting with pop stacks with a bypass, the priority of push over bypass can be a weakness when trying to modify to sort words.  When stack sorting, the patterns that must be avoided to be sortable are the same for permutations and words.  However, the same is not true for pop stack sorting where, in addition to avoiding $213$ and $312$, words must also avoid $1010$, as shown for Cayley words by Cerbai~\cite{Ce} and later expanded to all words in \cite{MSS}.  The same additional restriction of avoiding $1010$ for words applies to \textsf{PSB}.

If we have a regular word with the same number of copies of each letter, for example Stirling permutations, or if we want to check the count of each letter prior to sorting, we can remove the need to avoid $1010$ by prioritizing the bypass operation more.  It can be shown using techniques similar to those given in Section~\ref{sortable} that Algorithm \ref{queuesort2}, which we call \textsf{PSBW}$(k)$, gives the procedure for $k$ regular words with the smallest letter $0$.  This family of algorithms could also be expanded to keep track of individual repetitions to properly prioritize the bypass when sorting other types of words.

\begin{algorithm}
	$S:=\emptyset$\;
	$i:=1$\;
    $j:=0$\;
	\While{$i\leq kn$}
	{
		\If{$\pi_i \leq \lfloor \frac{j}{k} \rfloor$ }
		{
			\texttt{BYPASS}\;
             $j := j+1$;
		}
		\ElseIf{$\pi_i \leq \textnormal{\texttt{TOP}}(S)$}
		{
			\texttt{PUSH}\; 
		}
        \Else
		{
			\texttt{POP}\;
            $j := i-1$\; 
            \If{$\pi_i \leq \lfloor \frac{j}{k} \rfloor$ }
		      {
			     \texttt{BYPASS}\;
                $j := j+1$ \;
		      }
            \Else
            {
			     \texttt{PUSH}\;
            }
		}
		$i:=i+1$\;		
	}
	\texttt{POP}\;
	\caption{\textsf{PSBW}$(k)$ ($S$ is the pop stack; \texttt{TOP}(S) is the current top element of the pop stack; 
		$\pi=\pi_1 \cdots \pi_{kn}$ is the input $k$-regular word (for a permutation, $k=1$); $j$ is the number of entries in the output;  \texttt{TOP}(S) at the top of pop stack.}\label{queuesort2}
\end{algorithm}

\subsection{Conjecture on the enumeration of simple permutations sortable by two pop stacks in series}

We have already noted that the terms of the generating function for the sequence corresponding to the permutations of size $n$ sortable by $k=2$ pop stacks in parallel now appear in \cite{Sl} as A374165. However, there is also a potential nice closed form for the simple sortable permutations that we boldly put here as a conjecture:

\begin{conj}
	Let $a_n$ be the number of simple permutations of size $n$ that can be sorted by a machine consisting of two pop stacks in parallel where entries are allowed to bypass the pop stacks. Then $a_0 =a_1 =1$, $a_2 =2$, $a_n =F_{2n-5}-1$ if $n\geq 3$ is odd, and $a_n =F_{2n-5}$ if $n > 3$ is even (where $F_n$ is the $n$-th Fibonacci number).
\end{conj}

\end{document}